\documentclass[letterpaper]{article}

\usepackage[T1]{fontenc}
\usepackage[utf8]{inputenc}
\usepackage{geometry}
\usepackage{setspace}
\usepackage{graphicx}
\usepackage{pdfpages}
\usepackage{amsmath,amssymb,mathrsfs,amsfonts,mathtools}
\usepackage{bm}
\usepackage{subcaption}
\usepackage{xcolor}
\usepackage[flushleft]{threeparttable}
\usepackage{booktabs}
\usepackage{makecell}
\usepackage{multirow}
\usepackage[version=4]{mhchem}
\usepackage[hidelinks]{hyperref}
\usepackage[capitalise]{cleveref}
\usepackage[
    backend=biber,
    style=chem-acs,
    articletitle=true,
    doi=false,
    url=false,
    eprint=false,
    isbn=false
]{biblatex}
\usepackage{authblk}

\newcommand{\I}{\mathrm{i}}
\newcommand{\mc}{\mathcal}
\newcommand{\dd}{\mathop{}\!\mathrm{d}}
\newcommand{\dv}[2]{\frac{\mathrm{d}#1}{\mathrm{d}#2}}
\newcommand{\ket}[1]{\left|#1\right\rangle}
\newcommand{\bra}[1]{\left\langle#1\right|}

\newcommand{\mel}[3]{\left\langle#1\middle|#2\middle|#3\right\rangle}
\newcommand{\abs}[1]{\left|#1\right|}
\newcommand{\norm}[1]{\left\lVert#1\right\rVert}

\newcommand{\tnorm}[1]{{\left\vert\kern-0.25ex\left\vert\kern-0.25ex\left\vert#1\right\vert\kern-0.25ex\right\vert\kern-0.25ex\right\vert}}

\newcommand{\ud}{\,\mathrm{d}}

\newcommand{\HL}[1]{}
\newcommand{\LL}[1]{}
\newcommand{\rev}[1]{#1}
\makeatletter
\renewcommand*\env@matrix[1][\arraystretch]{%
        \edef\arraystretch{#1}%
        \hskip -\arraycolsep
        \let\@ifnextchar\new@ifnextchar
        \array{*\c@MaxMatrixCols c}}
\makeatother
\crefname{equation}{eq}{eqs}
\Crefname{equation}{Eq}{Eqs}
\crefname{figure}{Fig.}{Figs.}
\Crefname{figure}{Fig.}{Figs.}
\crefname{table}{Table}{Tables}
\Crefname{table}{Table}{Tables}

\author[1]{Hao-En Li}
\author[1,2]{Lin Lin*}
\affil[1]{Department of Mathematics, University of California, Berkeley, California 94720, United States}
\affil[2]{Applied Mathematics and Computational Research Division, Lawrence Berkeley National Laboratory, Berkeley, California 94720, United States}

\title{Dissipative Quantum Algorithms for Excited-State Quantum Chemistry}
\date{*Email: linlin@math.berkeley.edu}

\begin{document}

\maketitle

\begin{abstract}
 
Electronic excited states are central to a vast array of physical and chemical phenomena, yet accurate and efficient methods for preparing them on quantum devices remain challenging and comparatively underexplored. We introduce a general dissipative algorithm for selectively preparing \textit{ab initio} electronic excited states. The key idea is to recast excited-state preparation as an effective ground-state problem by suitably modifying the underlying Lindblad dynamics so that the target excited state becomes the unique steady state of a designed quantum channel. We develop three complementary strategies, tailored to different types of prior information about the excited state, such as symmetry and approximate energy. We demonstrate the effectiveness and versatility of these schemes through numerical simulations of atomic and molecular spectra, including valence excitations in prototypical planar conjugated molecules and transition-metal complexes. Taken together, these results provide a new pathway for advancing quantum simulation methods for realistic strongly correlated electronic systems.
\end{abstract}
 
\section{Introduction} \label{sec:Introduction}

Many important chemical processes, including photochemical reactivity~\cite{GonzalezLindh2020}, energy transfer in light-harvesting complexes~\cite{Scholes2017}, transition-metal photoredox catalysis~\cite{PrierRankicMacMillan2013}, and spectroscopic characterization~\cite{Kalsi2007}, require accurate descriptions of electronic excited states~\cite{ZhaoNeuscamman2019,PfauAxelrodSutterudEtAl2024}. 
On a fault-tolerant quantum computer, a natural candidate algorithm for preparing energy eigenstates is quantum phase estimation (QPE).
Given an initial state with sufficient overlap with the target eigenstate, QPE can project out the desired eigenstate and output the corresponding eigenvalue~\cite{AbramsLloyd1999, NielsenChuang2010}.
However, for chemically relevant excited states, such good initial states can be difficult to construct \cite{LeeLeeZhaiEtAl2023}. From the perspective of computational complexity, preparing the ground state of a local Hamiltonian is already QMA-hard in the worst case~\cite{KitaevShenVyalyi2002, KempeKitaevRegev2006, AharonovGottesmanIraniEtAl2009}, and excited-state preparation should be at least as hard.

\rev{For many proposed quantum algorithms for excited-state electronic structure problems, success likewise depends on the quality of constructed subspaces or initial states. This applies to methods such as quantum subspace expansion (QSE)\cite{McCleanKimchiSchwartzCarterEtAl2017}, quantum Krylov/filter diagonalization \cite{StairHuangEvangelista2020,CohnMottaParrish2021,CortesGray2022,EpperlyLinNakatsukasa2022}, quantum selected configuration interaction (QSCI)~\cite{KannoKohdaImaiEtAl2026}, post-Kitaev  phase-estimation methods~\cite{Somma2019,OBrienTarasinskiTerhal2019,LinTong2022,DingLin2023,DingLiLinNiEtAl2024}, to name a few.}
These considerations motivate the development of quantum algorithms that \emph{do not} rely on high-quality initial states, yet still make use of system-specific structure rather than treating the physical or chemical problem in a fully agnostic way.  
In this work, we focus on dissipative state preparation (DSP) schemes~\cite{KrausBuchlerDiehlEtAl2008,VerstraeteWolfCirac2009}.
In DSP, the quantum many-body system of interest is coupled to an engineered environment that induces dissipation, and the target state is encoded as the steady state of the resulting dynamics. Dissipative algorithms can be viewed as quantum generalizations of Markov chain Monte Carlo (MCMC) methods: by carefully designing the dissipative mechanism, incorporating system-specific input, and repeatedly applying the quantum channel that simulates the Lindblad dynamics, the system is driven toward the steady state of that channel~\cite{Lin2025}.  Compared to alternatives such as variational quantum algorithms (VQA)~\cite{TillyChenCaoEtAl2022}, adiabatic state preparation (ASP)~\cite{FarhiGoldstoneGutmannEtAl2000}, and quantum imaginary-time evolution (QITE)~\cite{McArdleJonesEndoEtAl2019}, dissipative schemes can avoid variational parameter optimization, full state tomography, and normalization factors that may grow rapidly with system size. Dissipative algorithms are also intrinsically robust to certain forms of noise due to the contractive nature of the dynamics.

In recent years, there has been significant progress in using dissipative dynamics to prepare thermal and ground states~\cite{RoyChalkerGornyiEtAl2020,Cubitt2023,MiMichailidisShabaniEtAl2024,RallWangWocjan2023,ChenKastoryanoBrandaoEtAl2025,ChenKastoryanoGilyen2023,DingLiLin2025,DingChenLin2024,LiZhanLin2025,ZhanDingHuhnEtAl2025,HahnSwekeDeshpandeEtAl2025,GilyenChenDoriguelloKastoryano2024,JiangIrani2024,LambertCirioLinEtAl2024,EderFinzgarBraunEtAl2025,MatthiesRudnerRoschEtAl2024,MotlaghZiniArrazolaEtAl2024,PerrinScoquartPavlovEtAl2025,HaganWiebe2025,LloydMichailidisMiEtAl2025}. 
For ground-state problems, the effectiveness of the dissipative protocol, using algorithmically designed Lindblad dynamics~\cite{DingChenLin2024} has been demonstrated for both spin systems~\cite{ZhanDingHuhnEtAl2025} and \textit{ab initio} electronic Hamiltonians~\cite{LiZhanLin2025}. These studies provide rigorous guarantees in certain parameter regimes, as well as numerical evidence of good performance beyond the provably controlled setting.

The goal of this work is to extend dissipative quantum dynamics to the preparation of electronic excited states on quantum computers.
Our main idea is to convert the excited-state preparation problem into an \emph{effective} ground-state problem by embedding the target excited state as the steady state of a modified dissipative evolution. Depending on the available prior information, the construction can be tailored in different ways.  We design three complementary strategies that exploit prior knowledge such as the symmetry sector and approximate energy of the target excited state, and we investigate how these choices influence algorithmic performance and resource requirements.

We validate the resulting dissipative excited-state framework in \textit{ab initio} electronic structure simulations. Our numerical experiments include atomic spectra of second-period atoms, and excitation spectra of small molecular systems.
For larger systems, we employ active-space representations \cite{SayfutyarovaSunChanEtAl2017,SayfutyarovaHammes-Schiffer2019} to examine $\pi$-$\pi^\ast$ excitations in benzene (C$_6$H$_6$) and $d$-$d$ transitions in ferrocene (Fe(C$_5$H$_5$)$_2$).
Across these examples, our methods achieve accurate excited-state energy estimates and capture key physical properties such as spin multiplicities and qualitative transition character.  These results demonstrate the promise of dissipative state preparation as a flexible and potentially scalable primitive for quantum simulation of excited-state chemistry.
 
\section{Results}\label{sec:results}
\subsection{Excited-state preparation via dissipative dynamics}\label{sec:excitedstates}
\begin{figure}
    \centering
    \includegraphics[width=1.00\textwidth]{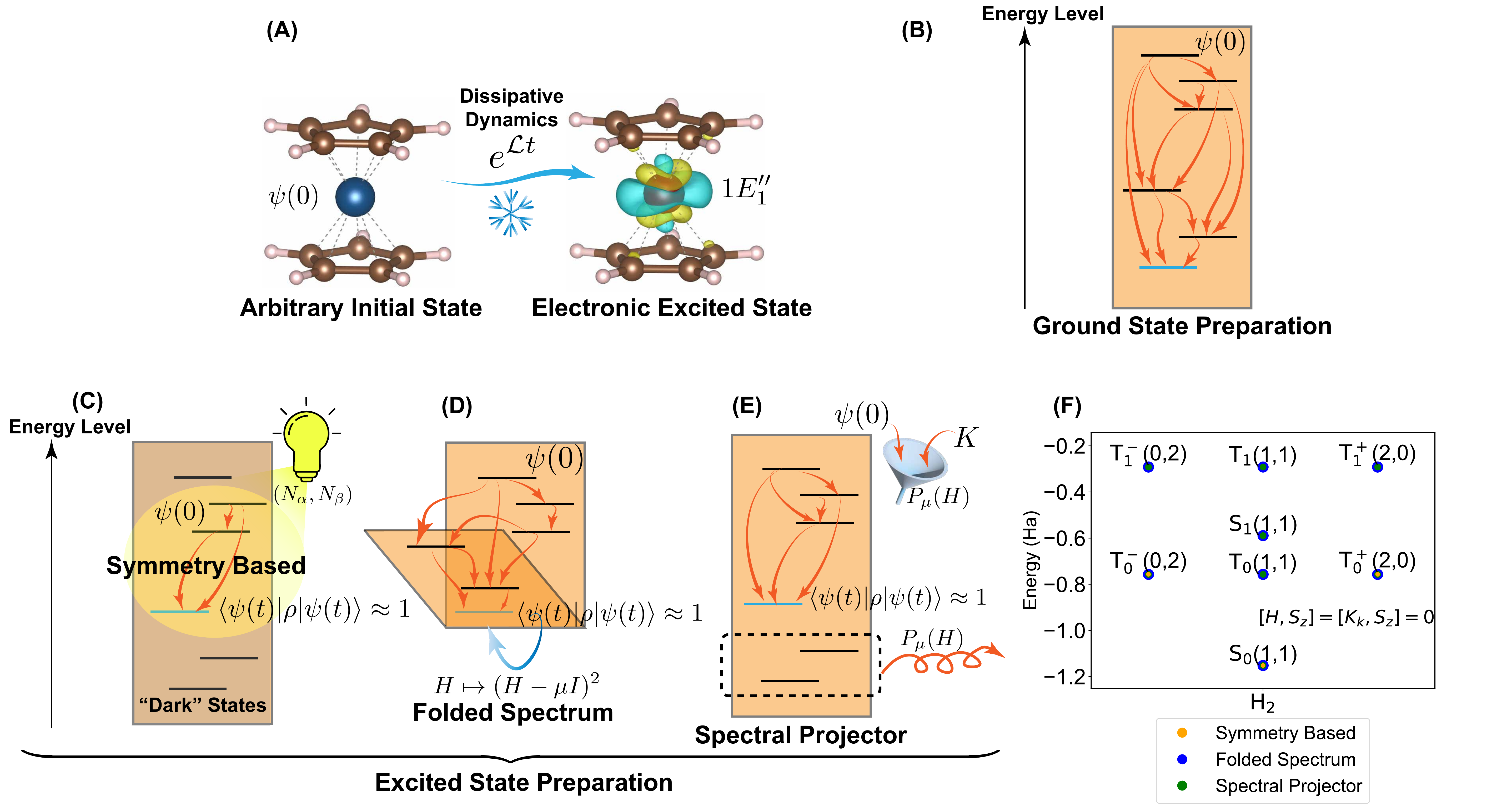}
    \caption{\raggedright\textbf{Schematic representation of the dissipative dynamics protocols for preparing excited states.} (\textbf{A}) The Lindblad dynamics is designed to drive the system from \emph{any} initial state to the desired excited state. The graph represents the isosurface of electron density differences (EDD) between the $1E_1''$ excited state and the ground state of ferrocene. (\textbf{B}) Mechanism of dissipative ground-state preparation. Every state has efficient energy-lowering transition pathways that connect it to the target ground state. (\textbf{C}) A symmetry-based approach for excited-state preparation. The initial state is restricted to a symmetry sector e.g. $(N_\alpha, N_\beta)$ matching the target excited state, so that lower-energy states outside this sector become ``dark'' and are avoided throughout the dissipative dynamics. The problem is then reduced to preparing the ground state within this symmetry sector. (\textbf{D}) ``Folded-spectrum'' approach for dissipative excited-state preparation. The transformation $H \mapsto (H - \mu I)^2$ folds the spectrum around a reference energy $\mu$, making the target excited state the effective ground state of the transformed Hamiltonian. (\textbf{E}) A spectral projector approach. We construct an operator $P_{\mu}(H)$ which projects out the energy eigenstates with energy below $\mu$, and then apply $P_\mu(H)$ to the initial state as well as the jump operators, effectively removing the possibility of decaying into states with energy below $\mu$.     (\textbf{F}) Energy levels and the symmetry sectors of the hydrogen molecule $\rm H_2$ in the 6-31G basis set. The $(N_\alpha,N_\beta)$ sectors of the low-lying states are indicated in the figure. The orange, blue, green markers represent that this target state can be prepared by the symmetry-based, folded-spectrum, and spectral projector methods, respectively.}
    \label{fig:excited_states}
\end{figure}

The key idea of the dissipative excited-state preparation framework is to engineer the system-environment coupling described by Lindblad dynamics, which can in principle drive the system to the electronic excited state of interest from any initial state, as illustrated in Fig.~\ref{fig:excited_states}a. This method builds upon the recently developed dissipative ground-state preparation protocols \cite{DingChenLin2024,ZhanDingHuhnEtAl2025,LiZhanLin2025, Lin2025}. Specifically, we consider the following state-preparation Lindblad dynamics with Hamiltonian $H$ and jump operators $\{K_k\}$:
\begin{equation}\label{eq:lindblad}
    \begin{aligned}
    \dv{\rho}{t} = -\mathrm{i}[H,\rho] + \sum_k \left(K_k \rho K_k^\dagger - \frac{1}{2}\{K_k^\dagger K_k, \rho\}\right).
    \end{aligned}
\end{equation}
The jump operators are constructed from a set $\mc A$ of coupling operators and a filter function $f$, which together determine the transitions between the eigenstates of $H$ induced by the dissipative dynamics (see Sections~\ref{sec:review} and \ref{sec:coupling_operators} in Methods). By construction, the desired excited state $\ket{\psi_\text{target}}$ is annihilated by every jump operator: 
\begin{equation}
    K_k \ket{\psi_\text{target}} =0,
\end{equation}
which hence ensures that $\ket{\psi_{\rm target}}$ is a steady state (or fixed point) of the dynamics. For example, for ground state preparation, the construction of the jump operators satisfies $\mel{\psi_i}{K_k}{\psi_j} = 0$ if $\lambda_i \ge \lambda_j$, where $\{\lambda_i, \ket{\psi_i}\} $ are the eigenpairs of $H$ and $\ket{\psi_0}$ is the ground state.  Thus, the dissipative dynamics only induces energy-lowering transitions (see Fig.~\ref{fig:excited_states}b). Similarly, if the coupling operators are designed such that every other eigenstate has efficient transition pathways that connect it to the target excited state $\ket{\psi_{\rm target}}$, then the system will converge to $\ket{\psi_{\rm target}}$ from \emph{any} initial state after a suitable time (called the \textit{mixing time}) of dissipative dynamics evolution.

For this purpose, we may modify the effective ordering of the eigenstates ``seen'' by the dissipative evolution, so that the desired excited eigenstate plays the role of a ground state within an appropriately chosen manifold. Below, we describe three such modifications,  which provide complementary ways to realize this effective ground-state picture for a wide class of excited states.

The first strategy is to exploit Hamiltonian symmetries. If both the initial state and the coupling operators are restricted to a chosen symmetry sector, then the dissipative evolution remains confined to that sector. When the target excited state is the lowest-energy state in this sector, its preparation reduces to a ground-state problem within the restricted subspace (Fig.~\ref{fig:excited_states}c). Concretely, we select symmetry-preserving coupling operators so that all states in other sectors are ``dark'' under the dynamics, and hence never populated. This is an effective and low-cost route to prepare excited states that differ from the ground state by, for example, particle-number or spin quantum numbers.

In many cases, the excited states of interest are in the same symmetry sector as the ground state. When an approximate energy $\mu$ of the target state is available, we can instead use an energy-targeting strategy. This assumption is well justified in practical quantum chemistry applications, where preliminary excited-state energies are typically obtained from low-cost electronic structure methods, and the goal is to refine these estimates and accurately prepare the corresponding excited states. Thus, the second strategy is the folded-spectrum method, which applies the transformation $H \mapsto (H - \mu I)^2$ so that the target becomes the ground state of the modified operator (Fig.~\ref{fig:excited_states}d).

However, the squaring of the Hamiltonian amplifies both the spectral radius and the inverse gap, which makes the underlying time evolutions significantly more expensive (see the resource estimation in Methods Section~\ref{sec:resource_estimates} and Supporting Information, Section S1). To reduce this cost while still exploiting approximate energy information, our third strategy is a spectral-projector approach: we construct an operator $P_\mu(H)$ that approximately projects out all eigenstates with energies below a reference $\mu$, and apply $P_\mu(H)$ to both the jump operators and the initial state. This spectral projector allows us to avoid simulating $(H - \mu I)^2$, which trades a more complex implementation of the jump operators for a reduction in overall cost. After this filtering, the dissipative dynamics act only within the subspace above $\mu$ and converge to the lowest-energy eigenstate in that subspace, i.e., the target excited state (Fig.~\ref{fig:excited_states}e).
We refer readers to Section~\ref{sec:excitedstates_long} in Methods for more details on these three protocols.

In most cases considered in this paper, we adopt the quadratic coupling operator set $\mc A = \mc S_{\rm II}$, or a reduced version $\mc A = \mc S_{\rm II}'$. This choice is motivated by our previous study \cite{LiZhanLin2025}, in which both theoretical and numerical analyses demonstrated that this generic class of coupling operators is effective for dissipative ground-state preparation in quantum chemistry applications. Further details regarding the construction of these coupling operators are provided in Methods Section~\ref{sec:coupling_operators}.

\subsection{Validation on hydrogen molecules}\label{sec:demonstration}

We validate the performance of the proposed protocols using the $\rm H_2$ molecule in the 6-31G basis set, and the equidistant $\rm H_4$ chain in the STO-3G basis set.  The bond lengths in this section are set to be $0.7$ \AA. All benchmark and reference quantum chemical calculations presented in this work were performed using the \texttt{PySCF} package \cite{SunBerkelbachBluntEtAl2018,SunZhangBanerjeeEtAl2020}.

 Without an external magnetic field, the first excited manifold of the hydrogen molecule systems consists of three degenerate triplet states, $\rm T_0$ and $\rm T_0^{\pm}$, with $(N_\alpha, N_\beta) = (1,1), (2,0)$ and $(0,2)$, respectively. The states $\rm T_0^\pm$ lie in different symmetry sectors from the ground state $\rm S_0$, so they can be prepared directly using the symmetry-based approach by restricting the initial state to the corresponding $(N_\alpha, N_\beta)$ sectors. In contrast, $\rm T_0$ shares the same symmetry sector as $\rm S_0$, and therefore should be prepared using either the folded-spectrum method or the spectral-projector method.

We evaluate the efficiency of these protocols by preparing the lowest triplet state $\rm T_0$ and its spin-adjacent partners $\rm T_0^{\pm}$ for both $\rm H_2$ and $\rm H_4$.  Fig.~\ref{fig:excited_states}f lists the low-lying energy levels and their $(N_\alpha, N_\beta)$ symmetry sectors for $\rm H_2$.
For $\rm T_0^{\pm}$, the symmetry-based protocol can be applied directly by initializing the system in the corresponding $(N_\alpha, N_\beta)$ sector. For $\rm T_0$, which shares the same sector as the ground state, both the folded-spectrum and spectral-projector protocols are used. 
The initial state is taken as the restricted or restricted open-shell Hartree--Fock (RHF or ROHF) ground state in the relevant spin sector.

\begin{figure}[t!]
       \includegraphics[width = 0.99\textwidth]{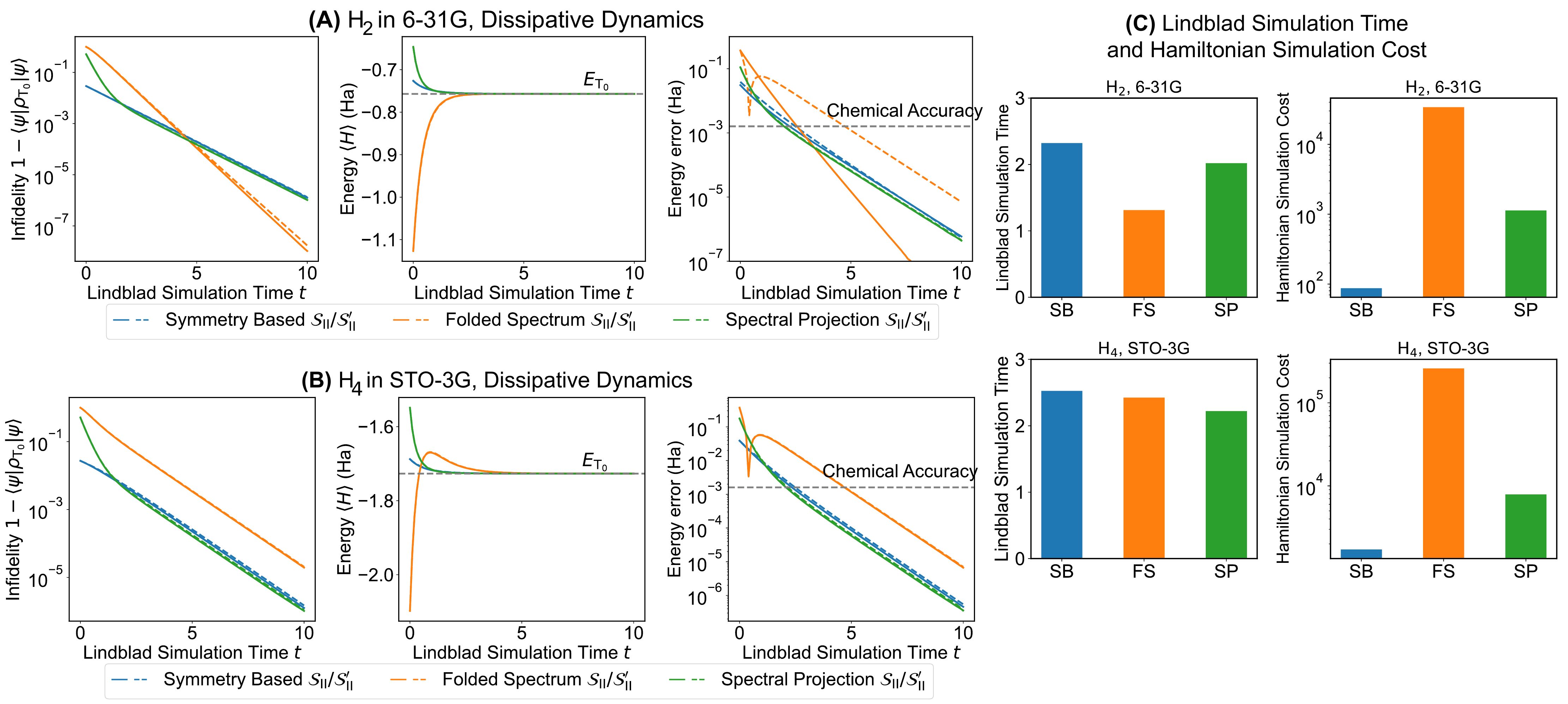}
    \caption{\raggedright\textbf{Performance of the three protocols for preparing the $\rm T_0$ and $\rm T_0^{\pm}$ states for $\rm H_2$ in 6-31G and $\rm H_4$ in STO-3G.} Convergence of the infidelity, energy expectation and energy error for (\textbf{A})  $\rm H_2$ in 6-31G and (\textbf{B}) $\rm H_4$ chain in STO-3G; (\textbf{C}) Lindblad simulation time and Hamiltonian simulation costs required to achieve chemical accuracy in energy for these three different excited-state preparation protocols, \rev{using the reduced coupling operator set $\mc S_{\rm II}'$ for the plotted estimates}.}\label{fig:performance} 
 
\end{figure}

As shown in Fig.~\ref{fig:performance}a and Fig.~\ref{fig:performance}b, all three protocols successfully prepare the target triplet states with high accuracy. Using the reduced coupling operator set $\mc S_{\rm II}'$ results in only a minor performance decrease compared to the full set $\mc S_{\rm II}$. We note that, despite the similar behavior in terms of the Lindblad simulation time, estimating the Hamiltonian-simulation cost reveals that the folded-spectrum approach is substantially more expensive than the other methods due to squaring the Hamiltonian, as shown in Fig.~\ref{fig:performance}c. A more detailed discussion of the resource estimation is provided in Methods Section~\ref{sec:resource_estimates} and Supporting Information, Section S1.

\subsection{Comparison with adiabatic state preparation and noise robustness}\label{sec:comparison}

Excited states can in principle be prepared via adiabatic state preparation (ASP), provided the target state remains separated from the rest of the spectrum by a non-zero energy gap throughout the evolution. However, the widely used Hartree--Fock (HF) Hamiltonian can lead to a challenge at the very start of the initialization (adiabatic parameter $s=0$). As detailed in Supporting Information, Section S4, the initial HF Hamiltonian often features degenerate excited states. Such degeneracy must be lifted, for example, via explicit electron-photon coupling as proposed in recent work \cite{BurtonFilip2025}, so that the ASP evolution can begin from a unique eigenstate.

In fact, the mean-field reference can already cause difficulties for ground-state ASP. For instance, during the bond dissociation
in hydrogen chains, the Coulson--Fischer (CF) point marks a mean-field instability, where the Restricted Hartree--Fock (RHF) solution transitions from a local minimum to a saddle point. As a result, the lowest mean-field excitation energy goes to zero, causing the adiabatic path gap to close, even though the true many-body spectral gap remains open. In contrast, DSP avoids these mean-field pitfalls and maintains accuracy across the potential energy surface (see Supporting Information, Section S5 for detailed discussion).

\begin{figure}[htbp]
    \centering 
    \includegraphics[width=0.99\textwidth]{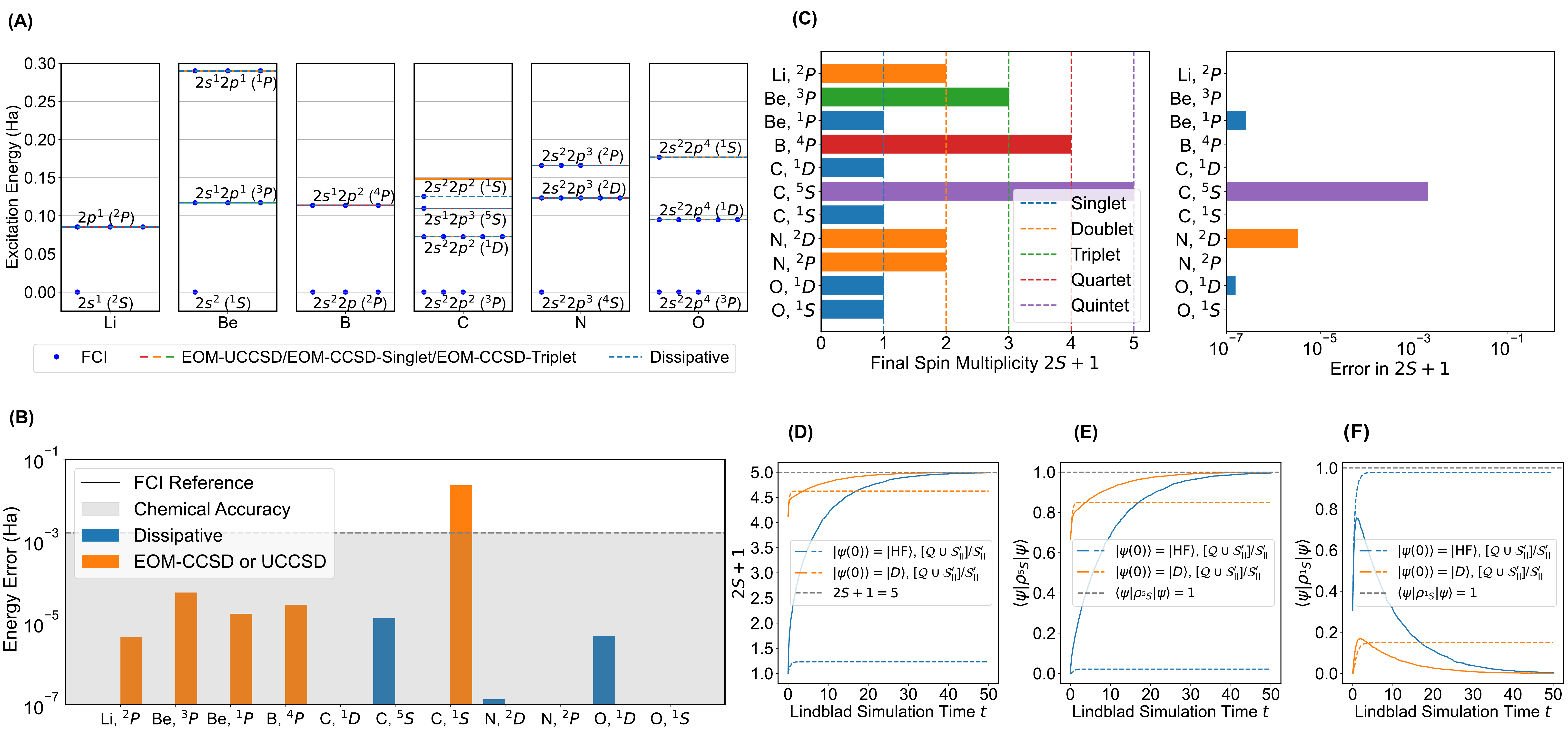}
    \caption{ \raggedright\textbf{Dissipative preparation of excited states in atomic systems.} (\textbf{A}) Energy spectrum of the second-period elements. The dashed lines represent the energy estimates obtained from the final states of the dissipative dynamics. The scatter points and the solid lines denote the FCI reference energies and the results from the EOM-CCSD/EOM-UCCSD methods, respectively. (\textbf{B}) Energy errors of the final states from the dissipative dynamics, EOM-CCSD, and EOM-UCCSD methods with respect to the FCI reference energies. (\textbf{C}) Spin multiplicities of the final states from the dissipative dynamics. The results show that the final states reproduce the correct spin multiplicities of the target excited states, indicating that they are highly accurate approximations to those states. (\textbf{D}-\textbf{F}) Dissipative preparation of the $2s^12p^3 (^5S)$ state of the carbon atom. The blue and orange lines represent the dynamics with initial state given by the HF ground state $\ket{\rm HF}$ and the high-spin initial state $\ket{D}$ (see Supporting Information, Section S9), respectively. The dashed lines denote the dissipative dynamics using only the reduced quadratic coupling operator set $\mc S_{\rm II}'$, while the solid lines correspond to the dynamics with the augmented coupling operator set \rev{$\mc Q\cup \mc S_{\rm II}'$}.
    (\textbf{D}) Spin multiplicity during the Lindblad dynamics. (\textbf{E}) The overlap with the $^5S$ state during the Lindblad dynamics.
     (\textbf{F}) The overlap with $^1S$ state during the Lindblad dynamics. }\label{fig:atomic_spectra}
\end{figure}

The dissipative state preparation framework can also be more robust against certain types of errors and noise that may arise in practical quantum devices. The ASP method may be more susceptible to errors that can accumulate during the adiabatic evolution, potentially leading to deviations from the desired target state. We refer readers to Supporting Information, Section S6 for demonstration of robustness of the dissipative state preparation against the depolarizing noise.  

\subsection{Applications to atomic spectra}\label{sec:atomicspectrum}

To investigate the effectiveness of dissipative excited-state preparation, we first employ the folded-spectrum method to compute the low-lying excitation spectra of second-period atoms from lithium (Li) to oxygen (O) using a minimal STO-3G basis set. The results are summarized in Fig.~\ref{fig:atomic_spectra}. Although the vertical excitations computed with the minimal basis set can only involve the $2s$ and $2p$ orbitals, they already capture rich physical information about the low-lying excited states of these atomic systems. As shown in Fig.~\ref{fig:atomic_spectra}a and Fig.~\ref{fig:atomic_spectra}b, the quantum states prepared via dissipative dynamics yield highly accurate estimates of the excited-state energies, with all errors well below the chemical accuracy threshold relative to the reference energies obtained by full configuration interaction (FCI) calculations. Comparisons with the energy estimates obtained from the dissipative dynamics with those from the equation-of-motion coupled-cluster with single and double excitations (EOM-CCSD) method \cite{StantonBartlett1993,Krylov2008} or the equation-of-motion unrestricted coupled-cluster with single and double excitations (EOM-UCCSD) method \cite{BarnesPeterssonMontgomeryEtAl2009} calculations also show that the dissipative dynamics can often achieve even higher accuracy than these widely used excited-state methods within a modest Lindblad simulation time. The spin multiplicities in Fig.~\ref{fig:atomic_spectra}c also match those of the target states, confirming that the prepared states faithfully reproduce the essential features of the true excited states.

The $2s^12p^3\ ({}^5S)$ state of the carbon atom is more challenging. With the coupling operator set $\mc S_{\rm II}'$, even if we choose $\mu$ to target the ${}^5S$ state, and the initial state to be the high-spin initial state $\ket{D}$ (see Supporting Information, Section S9) with a significant overlap with ${}^5S$, 
 the dissipative dynamics  converges to the \rev{$2s^2 2p^2\ ({}^1S)$} state rather than the desired ${}^5S$ state (blue and orange dashed lines in Fig.~\ref{fig:atomic_spectra}d-f). The fundamental issue is that $\mc S_{\rm II}'$ lacks the connectivity to transition between the disjoint configuration spaces of ${}^1S$ and ${}^5S$. Specifically, for any quadratic coupling operator $A\in \mc S_{\rm II}$, the transition matrix elements effectively vanish ($\abs{\mel{\psi_{^1S}}{A}{\psi_{^5S}}}<10^{-13}$, see Supporting Information, Section S8) due to spin selection rules. Since there are no other intermediate states between these two states, the Lindblad dynamics becomes non-ergodic, and the ${}^1S$ configuration acts as a spurious dark state, trapping the system in a local minimum.

To resolve this, we employ a subspace connectivity diagnostic which identifies the disconnected states, and then augment the coupling set to bridge the disconnected graph components, so that $\mc A = \mc S_{\rm II}' \cup \mc Q$ (see Methods Section~\ref{sec:coupling_operators}). This restores the connectivity required to exit the dark state. With this augmented set, the dynamics successfully converge to the correct ${}^5S$ state (blue and orange solid lines in Fig.~\ref{fig:atomic_spectra}d-f).

\subsection{Applications to molecular systems}\label{sec:molecularsystems}

\begin{figure}[htbp]
\centering
    \includegraphics[width=0.99\textwidth]{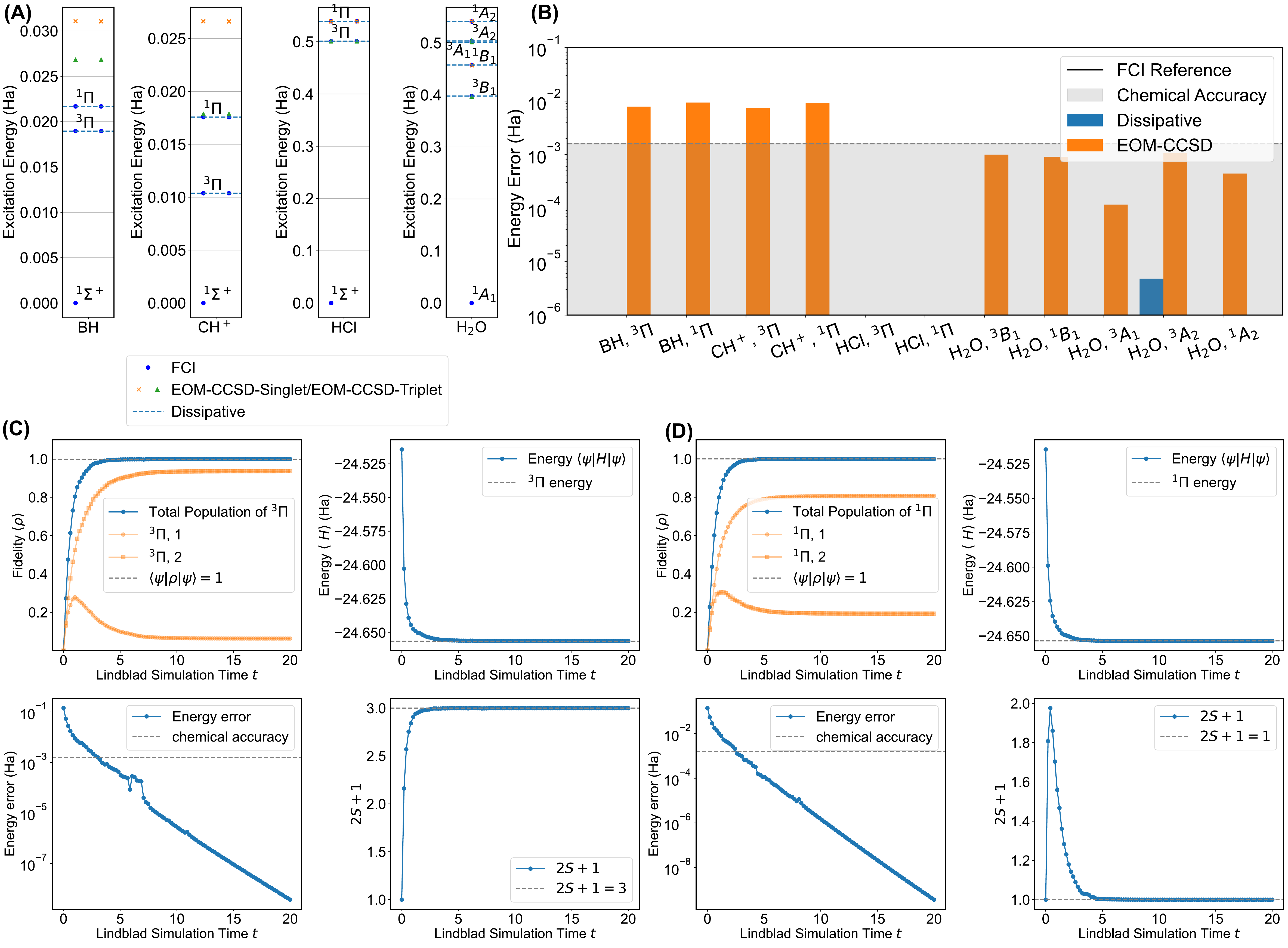}
    \caption{ \raggedright\textbf{Dissipative preparation of excited states in molecular systems.} (\textbf{A}) Energy spectrum of small molecules. The dashed lines represent the energy estimates obtained from the final states of the dissipative dynamics. The scatter points denote the FCI reference energies and the results from the EOM-CCSD and EOM-UCCSD methods. (\textbf{B}) Energy errors of the final states from the dissipative dynamics, EOM-CCSD, and EOM-UCCSD methods with respect to the FCI reference energies. (\textbf{C}) Convergence of the overlap, energy and spin multiplicities for $^3\Pi$ state of BH molecule. (\textbf{D}) Convergence of the overlap, energy and spin multiplicities for  $^1\Pi$ state of BH molecule. The orange lines in the plot for fidelity represent the overlap between the degenerate excited states, and the blue lines represent the total overlap with all the degenerate excited states at this energy level. Note that both $^3\Pi$ and $^1\Pi$ states are doubly degenerate.}\label{fig:molecular_spectra}
\end{figure}

We apply the dissipative excited-state preparation framework to small molecular systems, specifically $\rm BH$, $\rm CH^+$, $\rm H_2O$, and $\rm HCl$, using the STO-3G basis set. Here we employ the Monte-Carlo trajectory method to simulate the unraveled Lindblad dynamics, using $8\times 10^2$ trajectories for each simulation. The results are summarized in Fig.~\ref{fig:molecular_spectra}. 
We compare the energy estimates obtained from the dissipative dynamics with those from the EOM-CCSD method. For the $\rm BH$ and $\rm CH^+$ systems, we set the bond lengths to twice their respective equilibrium values ($2 \times 1.243$ {\AA} for $\rm BH$ and $2 \times 1.131$ {\AA} for $\rm CH^+$). At these stretched geometries, the ground states exhibit strong multireference character, which makes excited-state calculations particularly challenging for the EOM-CCSD method; this difficulty is already evident even in the minimal STO-3G basis set \cite{KowalskiPiecuch2001}. As shown in Fig.~\ref{fig:molecular_spectra}a and Fig.~\ref{fig:molecular_spectra}b, EOM-CCSD overestimates both the triplet $^3\Pi$ and singlet $^1\Pi$ excitation energies of $\rm BH$ and $\rm CH^+$ by nearly 10 mHa. In contrast, the dissipative dynamics with a stopping time of $T=20$ achieves accuracies well beyond chemical accuracy. For the $\rm H_2O$ and $\rm HCl$ systems at their equilibrium geometries, the ground and low-lying excited states are predominantly single-reference in nature, and EOM-CCSD therefore performs well, reaching chemical accuracy for most states. Nevertheless, the dissipative dynamics method often attains even higher accuracy, as illustrated in Fig.~\ref{fig:molecular_spectra}a and Fig.~\ref{fig:molecular_spectra}b.

We also investigate the convergence behavior of the dissipative dynamics for these molecular systems. As shown in Fig.~\ref{fig:molecular_spectra}c, the total overlap with all degenerate excited states at the target energy level, the energy expectation, and the spin multiplicity of the final states all converge efficiently to their desired values after the mixing time of the Lindblad dynamics. These results indicate that the dissipative dynamics can effectively prepare  the target excited states for molecular systems.

\subsection{Applications to the $\pi$-$\pi^\ast$ transitions in benzene and $d$-$d$ transitions in ferrocene}\label{sec:ferrocene}

We further apply the dissipative excited-state preparation framework to the $\pi$-$\pi^\ast$ transitions in benzene  (C$_6$H$_6$), and
the $d$-$d$ transitions in ferrocene (Fe($\mathrm{C_5H_5}$)$_2$). 

Benzene is a prototypical $\pi$-conjugated molecule, and its electronic excited states have been extensively studied both experimentally and theoretically \cite{LassettreSkerbeleDillonEtAl1968,NakashimaInoueSumitaniEtAl1980}. To render the dissipative state-preparation simulations computationally tractable, we employ a simplified (6e, 6o) model that focuses on the six valence $\pi$ orbitals, with three $\pi$ bonding orbitals and three $\pi^\ast$ antibonding orbitals. The active space is automatically constructed using the $\pi$-orbital space (PiOS) method \cite{SayfutyarovaHammes-Schiffer2019}. The selected active space orbitals are then optimized using the state-averaging complete active space self-consistent field (CASSCF) method. The optimized active space Hamiltonian is used in the dissipative dynamics simulations. We study the lowest six singlet excited states within this active space, which primarily correspond to $\pi$-$\pi^\ast$ excitations \cite{LassettreSkerbeleDillonEtAl1968,NakashimaInoueSumitaniEtAl1980}.

Ferrocene adopts a characteristic sandwich-like geometry in which a central iron atom is coordinated by two cyclopentadienyl (Cp) ligands. This structure gives rise to a rich manifold of electronic states, including low-lying valence excitations associated with $d$-$d$ transitions within the Fe $3d$ shell, as observed in early electronic absorption spectra studies \cite{ArmstrongSmithElderEtAl1967}. Prior quantum-chemistry investigations have shown that accurately describing these $d$-$d$ transitions requires careful treatment of the substantial multiconfigurational character of the relevant electronic states \cite{IshimuraHadaNakatsuji2002}. Here, we employ a simplified (10e, 7o) active-space model that focuses on the iron valence $3d$ orbitals generated using the atomic valence active space (AVAS) method. We analyze the lowest six triplet and six singlet excited states for ferrocene within the chosen active space. The excitations of ferrocene primarily correspond to $d$-$d$ transitions from the three nonbonding orbitals (dominated by $3d_{x^2-y^2}$, $3d_{xy}$, and $3d_{z^2}$) to the two antibonding orbitals (dominated by $3d_{xz}$ and $3d_{yz}$) of the iron center \cite{SayfutyarovaSunChanEtAl2017}. All of these excited states exhibit $E_1''$ or $E_2''$ symmetry and are doubly degenerate. For the details of active space construction for both benzene and ferrocene, we refer readers to Supporting Information, Section S7.

\begin{figure}[htbp]
    \centering
    \includegraphics[width=0.99\linewidth]{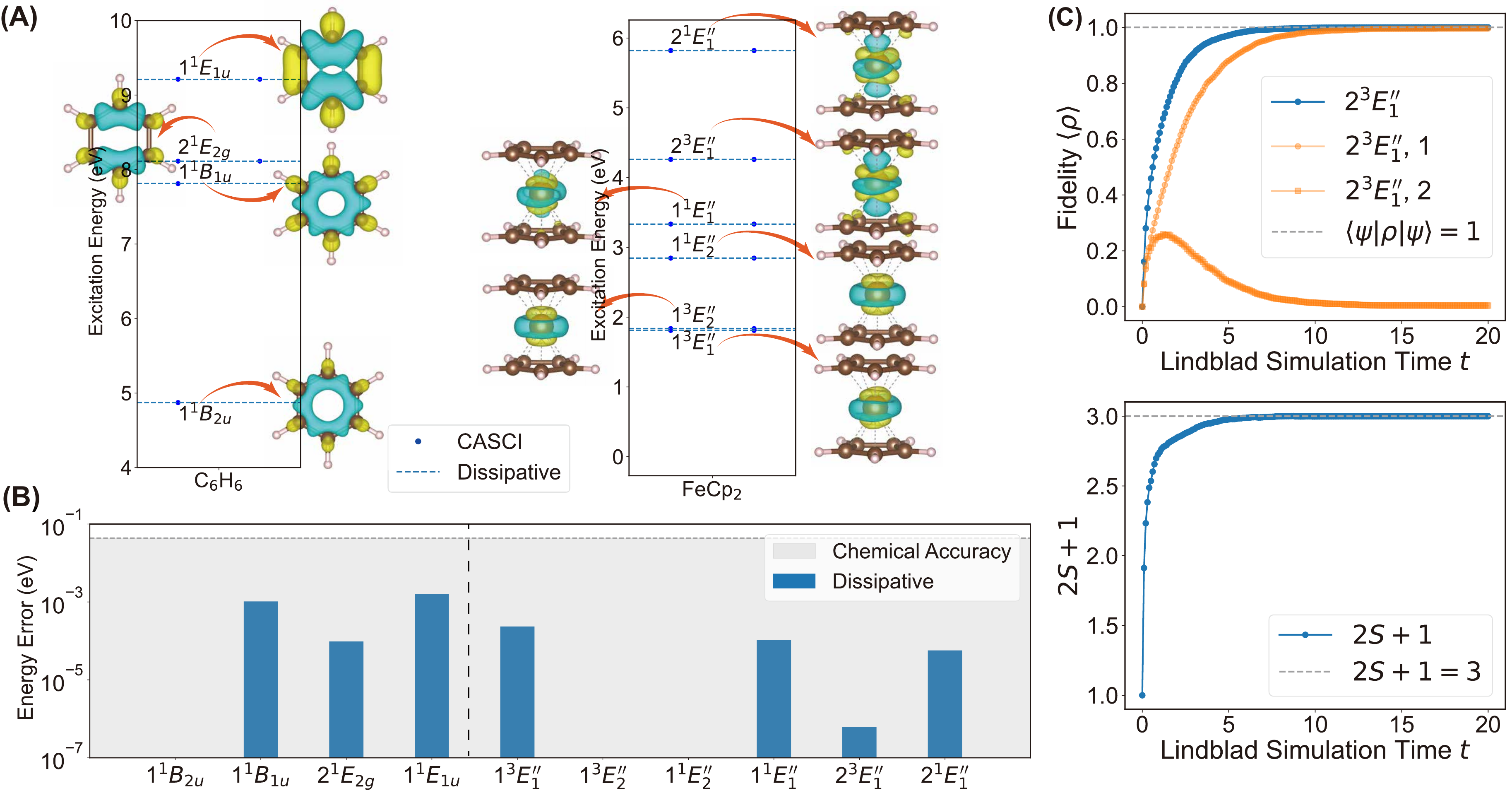}
    \caption{\raggedright\textbf{Dissipative preparation of the $\pi$-$\pi^\ast$ excited states in benzene and $d$-$d$ excited states in ferrocene.} (\textbf{A}) Energy spectrum of the low-lying excited states computed with the active space model. Dashed lines indicate the energy estimates obtained from the final states of the dissipative dynamics. The scatter points show the corresponding CASCI reference energies. Each state is accompanied by an isosurface plot of the electron density difference (EDD) between the excited and ground states (isosurface level = 0.0015 for $\rm C_6H_6$ and 0.005 for \rev{$\rm FeCp_2$}, identified from the converged density matrix), where yellow and blue indicate positive and negative values, respectively. The energies are given in eV.
(\textbf{B}) Energy deviations of the dissipatively prepared final states relative to the CASCI energies. The energies are in eV. (\textbf{C}) Convergence of the fidelity and spin multiplicity in the dissipative dynamics for preparation of the $2^3E_1''$ state. Note that the state possesses $E_1''$ symmetry and is doubly degenerate.  }\label{fig:ferrocene}
\end{figure}

For the numerical implementation, we apply the Monte Carlo trajectory based method with $2\times 10^3$ trajectories. 
The folded-spectrum method is employed with the reduced coupling operator set $\mathcal{S}_{\mathrm{II}}'$ for all but a few of the excited states. Owing to the presence of low-lying quintet states, the $1^1B_{1u}$, $2^1E_{2g}$, $1^1E_{1u}$ states (benzene) and
 $2^3 E_1''$, $2^{1}E_{1}''$ states (ferrocene) are treated using the augmented operator set $\mathcal{A} =   \mathcal{S}_{\mathrm{II}}'\cup \mathcal{Q} $, following the same strategy as in the carbon atom example. We set the initial state to be the HF ground state. 
The corresponding results are presented in Fig.~\ref{fig:ferrocene}, which show that the dissipative dynamics successfully prepares all of the excited states within the error threshold of chemical accuracy. Moreover, Fig.~\ref{fig:ferrocene}c illustrates the convergence of the fidelity and spin multiplicity during the dissipative dynamics used to prepare the ferrocene $2^3E_1''$ state, confirming that the target excited state is obtained both accurately and reliably. 

As shown in Table~\ref{table:casci}, the dissipative results agree well with experiment for all low-lying singlet and triplet excited states except $1^{1}E_{1u}$, $2^{1}E_{2g}$, $2^{3}E_{1}''$, and $2^{1}E_{1}''$. Because the active-space model includes only valence $\pi$-, $\pi^\ast$-, and $d$-orbitals, these states, which contain some Rydberg character, are distorted due to the absence of dynamical correlation with non-valence orbitals, leading to a systematic overestimation of their excitation energies \cite{RohmerVerillardWood1974,SayfutyarovaSunChanEtAl2017}. This behavior is also observed in several other electronic structure methods and can in general be mitigated by applying treatments that account for dynamical correlation \cite{HashimotoNakanoHirao1996,IshimuraHadaNakatsuji2002,FromagerKnechtJensen2013}.  
  Taken together, these results demonstrate that, for prototypical conjugated molecules and 
  transition-metal complexes with strong multiconfigurational character, the dissipative protocol can selectively prepare low-lying excited states with \rev{chemical accuracy relative to} complete active space configuration interaction (CASCI). The remaining discrepancies with experiment are largely attributable to limitations of the underlying CASCI model rather than to the dissipative dynamics itself. This strategy should likewise extend to the preparation of valence excited states in more complex molecular and materials systems, provided that an appropriately correlated Hamiltonian is available, for example from some \textit{ab initio} downfolding procedure \cite{ChanglaniZhengWagner2015,ChangvanLoonEskridgeEtAl2024,deSousaSayfutyarova2025}.

\begin{table}[!htbp]
    \centering
    \caption{\raggedright\textbf{Results for the lowest singlet and triplet excited states.} We compare the dissipative state preparation for the active space model with computational and experimental results. Energies are given in eV.}\label{table:casci}\begin{threeparttable}
    \begin{tabular}{ccccc}
    \toprule
    State & Dissipative & CASCI& CASSCF&  Experimental\\
    \midrule
$1^1 B_{2u}$ & $4.87$ & - &$4.97$$^\text{b}$, $4.82$$^\text{c}$ & $4.90$$^\text{d}$ \\
$1^1 B_{1u}$ & $7.81$ & - & $7.85$$^\text{b}$, $7.91$$^\text{c}$ & $6.20$$^\text{d}$\\
$2^1 E_{2g}$ & $8.11$ & - &$8.11$$^\text{b}$, $8.01$$^\text{c}$ & $7.80$$^\text{e}$\\
$1^1 E_{1u}$ & $9.21$ & - 
&$9.29$$^\text{b,c}$ & $6.95$$^\text{d}$\\
\hline
     $1^3 E_1''$ & $1.81$ & $1.81$$^\text{a}$     &-   & $1.74$$^\text{f}$ \\
     $1^3E_2''$ &$1.84$     & $1.84$$^\text{a}$  &- & $2.05$$^\text{f}$\\
     $1^1E_2''$ & $2.85$ &  $2.85$$^\text{a}$ &- & $2.8$$^\text{f}$, $2.7$$^\text{g}$\\
      $1^1E_1''$ &$3.33$&$3.33$$^\text{a}$ &-  & $2.81$$^\text{f}$, $2.98$$^\text{g}$ \\
      $2^3E_1''$ & $4.26$ & $4.26$$^\text{a}$  &- & $2.29$--$2.34$$^\text{f}$\\
       $2^1E_1''$ & $5.82$ & $5.82$$^\text{a}$ &-  & $3.8$$^\text{f,g}$\\
       \bottomrule\\

    \end{tabular}
    \begin{tablenotes}
\item $^\text{a}$ Ref. \cite{SayfutyarovaSunChanEtAl2017}, $^\text{b}$ Ref. \cite{RoosAnderssonFulscher1992}, $^\text{c}$ Ref. \cite{HashimotoNakanoHirao1996}, $^\text{d}$ Ref. \cite{LassettreSkerbeleDillonEtAl1968}, $^\text{e}$ Ref. \cite{NakashimaInoueSumitaniEtAl1980} $^\text{f}$ Ref. \cite{ArmstrongSmithElderEtAl1967}, $^\text{g}$ Ref. \cite{GraySohnHendrickson1971}. In the CASSCF calculations for benzene in the literature \cite{RoosAnderssonFulscher1992,HashimotoNakanoHirao1996}, the (6e, 6o) active space is manually selected, and the slight differences in the reported excitation energies may arise from different choices of basis sets and geometries \cite{SayfutyarovaHammes-Schiffer2019}.
    \end{tablenotes}
    \label{tab:placeholder}
    \end{threeparttable}
\end{table}

\section{Methods}\label{sec:methods}
\subsection{Overview of dissipative ground state preparation}\label{sec:review}

In this section, we briefly review the dissipative ground state preparation framework proposed in \cite{DingChenLin2024,LiZhanLin2025,ZhanDingHuhnEtAl2025}.
In this paper, we consider the second-quantized electronic structure Hamiltonian in a given finite basis set with $L$ (spatial) molecular orbitals (MOs)  
 \begin{equation}\label{eq:electronic_hamiltonian}
    \begin{aligned}
            H &= \sum_{p,q=1}^{L} h_{pq} \sum_{\sigma\in\{\alpha,\beta\}} c_{p\sigma}^\dagger c_{q\sigma}  + \frac{1}{2}\sum_{p,q,r,s=1}^{L} V_{pqrs} \sum_{\sigma,\tau\in\{\alpha,\beta\}} c_{p\sigma}^\dagger c_{q\tau}^\dagger c_{s\tau} c_{r\sigma},
    \end{aligned}
\end{equation}
where $c_{p\sigma}^\dagger$ and $c_{p\sigma}$ are the fermionic creation and annihilation operators for the spin-orbital indexed by $(p,\sigma)$, with $p$ being the spatial MO index and $\sigma\in\{\alpha,\beta\}$ denoting the spin index. The MOs are defined according to the canonical choice, such that the Fermi vacuum corresponds to the HF ground state computed in the given basis set. The one-electron integrals $h_{pq}$ and the two-electron integrals $V_{pqrs}$ can be obtained by rotating the atomic orbital (AO) integrals using the MO coefficients from a HF calculation. 

The jump operators are constructed as
\begin{equation}
    K_k = \sum_{i,j} \hat f(\lambda_i - \lambda_j) \mel{\psi_i}{A_k}{\psi_j} \ket{\psi_i}\bra{\psi_j},
\end{equation}
where $\mc A = \{A_k\}$ is the set of generic coupling operators that need to be carefully designed. In general, $A_k$ is required to be local to ensure efficient implementation of the jump operators on quantum devices.
The filter function $\hat f$ is chosen to have support only on the negative real axis, that is $\hat{f}(\lambda_i-\lambda_j)=0$ for any $\lambda_i \ge \lambda_j$. This construction ensures that the Lindblad dynamics only induces energy-decreasing transitions, i.e., transitions from higher-energy states to lower-energy states. As a result, the system will eventually converge to the ground state of $H$ regardless of the initial state.

The construction of the jump operator does not require explicitly diagonalizing $H$. Instead we can use the time domain representation:
\begin{equation}\label{eq:jump_time}
    K_k = \int_\mathbb{R} f(s) e^{\mathrm{i}Hs} A_k e^{-\mathrm{i}Hs} \ud s,
\end{equation}
where $f(s) = \frac{1}{2\pi} \int_\mathbb{R} \hat f(\omega) e^{-\mathrm{i}\omega s} \ud \omega$ is the inverse Fourier transform of $\hat f$. 
 
The frequency domain filter function $\hat f(\omega)$ is required to have support only on negative real axis. One specific choice for $\hat f$ is \cite{DingChenLin2024}
\begin{equation}
    \hat f(\omega) = \frac12 \left[\text{erf}\left(\frac{\omega+a}{\delta_a}\right) -\text{erf}\left(\frac{\omega+b}{\delta_b}
    \right)\right]
\end{equation}
which can be viewed as a smooth approximation to a characteristic function with support on $[-2\norm{H},-\Delta_H]$. The $\text{erf} $ function is $\text{erf}(\omega):=\frac2{\sqrt \pi} \int_0^{\omega}e^{-t^2}\dd t  $. The parameters $a$ and $b$ are chosen as the energy cutoff $a>2\norm{H}$ and the spectral gap of the Hamiltonian $b = \Delta_H$, respectively. The parameters $\delta_a$ and $\delta_b$ are chosen to be on the same order of $a$ and $b$ respectively. The inverse Fourier transform of $\hat f$ here can be analytically expressed as
\begin{equation}
    f(s) = \frac{1}{2\pi \I s}   \left(\exp(\mathrm{i}as-\delta_a^2s^2/4)-\exp(\mathrm{i}bs-\delta_b^2s^2/4)\right)
\end{equation}
with $f(0) =\frac{a-b}{2\pi}$ understood by taking the limit $s\to 0$. The time domain filter function $f(s)$ is approximately supported on an interval of length $S_s =\Theta(\frac1{\Delta_H})$, which enables efficient truncation of the infinite integral and numerical quadrature for quantum implementation of the jump operators
\begin{equation}\label{eq:Kk_quadrature}
    K_k \approx \int_{\abs{s}\le S_s}f(s) A_k(s)\dd s \approx \sum_{j=-M}^M w_j f(s_j) A_k(s_j)
\end{equation}
where $M$ is the number of 
quadrature nodes.

\subsection{Choice of coupling operators}\label{sec:coupling_operators}

We now discuss the choice of coupling operators $\mc A = \{A_k\}$ used to construct the jump operators in Eq.~\ref{eq:jump_time}.

Following our previous work \cite{LiZhanLin2025}, we choose the coupling operators from a set of Hermitian one-body operators that preserve $(N_\alpha,N_\beta)$ symmetry
\begin{equation}\label{eq:hermitian_ops}
\begin{aligned}
\mc S_{\rm II} &= \{c_{i\alpha}^\dagger c_{j\alpha} + c_{j\alpha}^\dagger c_{i\alpha} \}_{1\le i < j \le L}   \cup \{c_{i\beta}^\dagger c_{j\beta} + c_{j\beta}^\dagger c_{i\beta} \}_{1\le i < j \le L}.
\end{aligned}
\end{equation}
 
This choice of ``bulk'' coupling operators is defined over all pairs of molecular orbitals, yielding a set of $\mathcal{O}(L^2)$ operators. To reduce the cost of implementation, a more practical alternative is the reduced set $\mc S_{\rm II}'$, which retains only couplings between nearest- and next-nearest-neighbor molecular orbitals in energy level ordering:
\begin{equation}\label{eq:hermitian_ops_reduced}
        \begin{aligned}
    \mc S_{\rm II}' &= \{c_{i\alpha}^\dagger c_{j\alpha} + c_{j\alpha}^\dagger c_{i\alpha} \}_{|i-j| \le 2, 1\le i,j \le L}  \cup \{c_{i\beta}^\dagger c_{j\beta} + c_{j\beta}^\dagger c_{i\beta} \}_{|i-j| \le 2, 1\le i,j \le L}.
\end{aligned}
\end{equation}
This set contains only $\mc O(L)$ operators. As observed in \cite{LiZhanLin2025}, and will also be further demonstrated later in this paper, the numerical convergence behavior using the reduced set $\mc S_{\rm II}'$ is comparable to that using the full bulk set $\mc S_{\rm II}$ for ground state and excited state preparation. Furthermore, it is straightforward to verify that both $\mc S_{\rm II}$ and its reduced subset $\mc S_{\rm II}'$ preserve particle-number symmetry, making them suitable choices for preparing excited states within a fixed $(N_\alpha, N_\beta)$ sector.

 The success of dissipative dynamics does not require that every state be directly connected to every other state via a single jump operator, a condition that is expected to be challenging for the quasi-local jump operators constructed from local coupling operators. Rather, the method remains effective as long as there exists a viable transition path with sufficiently high probability. This condition is easily met in high-energy regimes where the density of states is high and numerous transition paths are available. However, ergodicity can be hampered between low-energy states separated by few or no intermediate states, which is a situation where transitions become ``dark''.

To address this, we can systematically augment the set of coupling operators using the following approach. We begin by defining a minimal test subspace, $\mathcal{P}$, spanned by a few physically relevant determinants, such as those within a CASCI space or a set of Hartree--Fock solutions and their local excitations. Precise knowledge of the exact eigenstates is not required; often, their symmetry properties suffice. For any state $\ket{\psi_i} \in \mathcal{P}$, we compute the effective transition rate to an approximate target state, $\ket{\psi_{\text{target}}}$, as:
\begin{equation}
    \Gamma_{i} = \sum_k \sum_{l=1}^{\ell} |\langle \psi_{\text{target}} | (K_k)^l | \psi_i \rangle|^2.
\end{equation}
Here, we allow for path lengths of up to $\ell$ jumps (where $\ell$ is typically small). If necessary, this quantity can be evaluated on a quantum computer using the SWAP test and block-encodings of the jump operator $K_k$.

For the specific case of the carbon atom (see details in Supporting Information, Section S8 with $\ell=1$), we expand beyond quadratic operators to include a selection of quartic operators.\rev{We provide a concrete demonstration of this diagonstic approach in Supporting Information, Section S8.} We define the augmented set as:
\begin{equation}\label{eq:augmented_ops}
    \mathcal{A} = \mathcal{Q} \cup \mathcal{S}_{\rm II}',
\end{equation}
where $\mathcal{Q}$ contains six additional quartic operators that preserve $(N_\alpha, N_\beta)$ symmetry:
\begin{equation}
    \begin{aligned}
    \mathcal{Q} = \{&c_{2\alpha}^\dag c_{3\alpha}^\dag c_{4\alpha} c_{5\alpha} + \text{h.c.},\quad c_{2\alpha}^\dag c_{3\alpha} c_{4\alpha} c_{5\alpha}^\dag + \text{h.c.}, \\
    &c_{2\beta}^\dag c_{3\beta}^\dag c_{4\beta} c_{5\beta} + \text{h.c.}, \quad
     c_{2\beta}^\dag c_{3\beta} c_{4\beta} c_{5\beta}^\dag + \text{h.c.},\\
    &c_{2\alpha}^\dag c_{3\alpha} c_{4\beta}^\dag c_{5\beta} + \text{h.c.}, \quad c_{2\alpha}^\dag c_{3\alpha} c_{4\beta} c_{5\beta}^\dag + \text{h.c.}\}.
    \end{aligned}
\end{equation}

\subsection{Implementation of excited-state preparation protocols}\label{sec:excitedstates_long}

In this section, we provide more details on the three strategies for excited-state preparation introduced in Section~\ref{sec:excitedstates}.
\subsubsection{Symmetry-based method} 
In the symmetry-based approach, we note that if the coupling operator $A\in \mc A$ is designed to preserve this symmetry, then the resulting jump operator will also preserve the symmetry, because of the construction of $K$ in Eq.~\ref{eq:jump_time}. Therefore, the Lindblad dynamics will be confined to this symmetry sector. 
In quantum chemistry applications, typical symmetries include point-group symmetry, particle-number symmetry, and spin symmetry. A particularly important example is the particle-number symmetry, characterized by the numbers of spin-up and spin-down electrons $(N_\alpha, N_\beta)$. This strategy has been widely and successfully employed, especially in quantum Monte Carlo methods \cite{FoulkesHoodNeeds1999,PurwantoZhangKrakauer2009,ShiZhang2013,MahajanSharmaZhangEtAl2025}. The key observation is that $[H,\hat S_z] = 0$, consequently, the full Fock space $\mc F$ decomposes into symmetry sectors labeled by $(N_\alpha, N_\beta)$, with $N_\alpha + N_\beta = N_{\mathrm{e}}$ the total number of electrons.

\subsubsection{Folded spectrum method }\label{sec:folded_spectrum}

As discussed in the main text, symmetry alone may be insufficient to isolate the desired excited state from the ground state and other low-lying states, and the relevant symmetry labels may be difficult to determine \emph{a priori}. Moreover, in practical settings it can be challenging to construct coupling operators $A$ that commute with all symmetry generators, which can render too many excited states ``dark,'' causing the Lindblad dynamics to become trapped and fail to converge to the intended excited state.
The ``folded-spectrum'' method \cite{WangZunger1994a,WangZunger1994b,CadiTaziThom2024} provides an alternative when an approximate energy $\mu$ of the target excited state is available, which transforms the original Hamiltonian $H$ into a new shifted-and-squared Hamiltonian $(H - \mu I)^2$, where $\mu$ is a reference energy close to the target excited state energy.
The jump operator takes the form
\begin{equation}\label{eq:folded_jump}
    K_k = \int_{\mathbb{R}} f(s)\, e^{\mathrm{i}(H - \mu I)^2 s}\, A_k\, e^{-\mathrm{i}(H - \mu I)^2 s} \,\mathrm{d}s .
\end{equation}

This transformation effectively folds the energy spectrum around $\mu$, making the target excited state the new ground state of the transformed Hamiltonian. It is also noted that as long as the reference energy $\mu$ is chosen such that
\begin{equation}
   \frac{\lambda_{\text{target}}+\lambda_<}{2}< \mu < \frac{\lambda_>+\lambda_{\text{target}}}{2} ,
\end{equation}
where $\lambda_>$ and $\lambda_<$ denote the eigenvalues of $H$ immediately above and below the target excited state energy $\lambda_{\rm target}$, respectively, 
the target excited state $\ket{\psi_{\rm target}}$ with energy $\lambda_{\rm target}$ will be the effective ground state of the folded-spectrum Hamiltonian $(H - \mu I)^2$. From this, we can see that the reference energy $\mu$ does not need to be extremely accurate and both the energy refinement and the subsequent state-preparation procedure remain robust under modest inaccuracies.

For the construction of the jump operators in the folded-spectrum approach for excited-state preparation~Eq.~\ref{eq:folded_jump}, it is required to implement the Heisenberg evolution governed by the folded-spectrum Hamiltonian \( (H - \mu I)^2 \). The function \( g(x) = e^{\I x^2} = \cos(x^2) + \I \sin(x^2) \) can be realized using quantum singular value transformation (QSVT) techniques, either by decomposing it into its even and odd components~\cite{GilyenSuLowEtAl2019,MartynRossiTanEtAl2021} or, more directly, through a single QSVT circuit~\cite{DongWhaleyLin2022}. 
\subsubsection{Spectral projector method }\label{sec:projector}

To convert the excited-state preparation task into a ground-state one, we can alternatively introduce a spectral projector that removes all eigenstates with energies below a chosen reference $\mu$:
\begin{equation}
K\mapsto P_\mu(H) K,
\end{equation}
Specifically, 
\begin{equation}
\begin{aligned}
    K 
    &= P_\mu(H) 
       \int_{\mathbb{R}} f(s)\, e^{\mathrm{i}Hs}\, A\, e^{-\mathrm{i}Hs}\, \mathrm{d}s \\[4pt]
    &= \sum_{\lambda_i,\lambda_j \in \mathrm{Spec}(H)\cap [\mu ,\infty)}
       \hat f(\lambda_i - \lambda_j) 
       \langle \psi_i | A | \psi_j \rangle 
       |\psi_i\rangle\langle \psi_j | ,
\end{aligned}
\end{equation}
where $P_\mu(H)$ is the orthogonal projector onto the subspace with eigenvalues above the cutoff $\mu$. This construction removes all energy-lowering transitions into states with energies below $\mu$, thereby restricting the dissipative dynamics to the manifold of eigenstates above the target energy. 
However, this modification also makes all states below $\mu$ steady states of the Lindblad dynamics. Therefore, we also need to apply this $P_\mu (H)$ to the initial state to eliminate the components below $\mu$. 

The spectral projector $P_\mu(H)$ can be written in terms of a filter function
 $g_\mu$ whose Fourier transform $\hat{g}_\mu$ has support only on $[\mu , +\infty)$:
\begin{equation}\label{eq:projector}
    P_\mu(H) = \hat{g}_\mu(H)  = \int_\mathbb{R} g_\mu(s) e^{\mathrm{i}Hs} \ud s,
\end{equation}
where $\text{supp}{\hat{g}_\mu} \subseteq [\mu , +\infty)$. The operator \( P_\mu(H) \) can be directly constructed using QSVT by choosing \( \hat{g}_\mu(\omega) \) in Eq.~\ref{eq:projector} to be, for example, a polynomial approximation to a step function. Alternatively, one may employ Hamiltonian simulation based on the Fourier representation with a suitably chosen filter function \( g_\mu \) in Eq.~\ref{eq:projector}. The latter approach ultimately leads to a quadrature approximation of a double integral, combined with the quadrature rule in Eq.~\ref{eq:Kk_quadrature} used for constructing the jump operators. 
Although the spectral projector method requires an additional step of implementing \( P_\mu(H) \), it avoids the squaring of the Hamiltonian inherent in the folded-spectrum method, which can lead to significant savings in implementation cost, as discussed in Section~\ref{sec:resource_estimates}.

\subsection{Resource estimates}\label{sec:resource_estimates}

In this section, we provide resource estimates for implementing the three excited-state preparation protocols on quantum devices. For dissipative state-preparation schemes based on Lindblad dynamics, the total cost can be divided into two components: 
(1) the cost $C$ of simulating the Lindblad dynamics per unit time, including the Hamiltonian-simulation cost associated with constructing the Lindbladian, in particular the jump operators; and (2) the required Lindblad simulation time $T$ needed to reach the target accuracy.
Therefore, the end-to-end cost of the protocol can be estimated by multiplying the two parts together
\begin{equation}
    \text{End-to-end cost} = C\times T.
\end{equation}

According to \cite{DingChenLin2024}, the Hamiltonian simulation cost for constructing a jump operator $K$ is $\widetilde{\mathcal O}(\norm{H} / \Delta_H)$. For the symmetry-based method, we consider the effective Hamiltonian restricted to the symmetry sector $\mathcal F_{(N_\alpha, N_\beta)}$. In this setting, the effective spectral radius $\|H_{\mathcal F_{(N_\alpha, N_\beta)}}\|$ is typically smaller than the original spectral radius $\norm{H}$, and the effective Hamiltonian gap $\Delta_{H|_{\mc F_{(N_\alpha, N_\beta)}}}$ is typically larger than the original $\Delta_H$. Therefore, when applicable, the symmetry-based method offers the most favorable cost scaling.

In the folded-spectrum method Eq.~\ref{eq:folded_jump}, the spectral radius of $(H - \mu I)^2$ scales as $\widetilde{\mathcal O}(\|H\|^2)$, while the corresponding spectral gap scales as $\widetilde{\mathcal O}(\Delta_H^2)$. Consequently, the Hamiltonian simulation cost for constructing the jump operators increases from $\widetilde{\mathcal O}(\|H\| / \Delta_H)$ for the original Hamiltonian $H$ to $\widetilde{\mathcal O}(\|H\|^2 / \Delta_H^2)$ for the folded Hamiltonian.

For the spectral projector method, according to Eq.~\ref{eq:projector}, the cost of implementing the polynomial approximation $\hat g_\mu(H)$ to the step function at $\mu$ using QSVT is $\widetilde{\mathcal O}(\|H\| / \Delta_H)$ \cite{GilyenSuLowEtAl2019, MartynRossiTanEtAl2021}. In the Fourier representation, the time-domain filter function $g_\mu(s)$ is approximately supported on an interval of size $\mathcal O(1 / \Delta_H)$, so the cost of implementing $P_\mu(H)$ using Fourier filtering is also $\widetilde{\mathcal O}(\|H\| / \Delta_H)$. Therefore, the total cost for implementing $P_\mu(H) K_k$ is $\widetilde{\mathcal O}(\|H\| / \Delta_H)$. After constructing the Lindbladian $\mathcal L$, define $\|\mathcal L\|_{\mathrm{be}} := \|H\| + \tfrac{1}{2} \sum_k \|K_k\|^2$. Using the algorithm in \cite{LiWang2023}, the query complexity for simulating the Lindblad dynamics up to time $t$ is $\widetilde{\mathcal O}( t \, \|\mathcal L\|_{\mathrm{be}} )$.

Based on the above analysis, to compare the total Hamiltonian simulation cost for different excited-state preparation protocols, we use the following estimate. For a Lindblad equation with effective Hamiltonian $H$, spectral gap $\Delta_H$, and jump operators $\{K_k\}$, the total Hamiltonian simulation cost required to reach chemical accuracy is  
\begin{equation}
    T \cdot C_K \cdot \left( \sum_k \|K_k\|^2 \right),
\end{equation}
where $T$ is the Lindbladian simulation time needed to achieve the desired accuracy, and $C_K$ is the Hamiltonian simulation cost for constructing each jump operator $K$ as analyzed above.

Using this estimate, we compare the three protocols as shown in Fig.~\ref{fig:performance}c. The Lindbladian simulation time $T$ and the norm of the jump operators $\sum_k \|K_k\|^2$ are comparable across all three protocols. The main discrepancy in the end-to-end cost arises from the factor $\|H\| / \Delta_H$, making the folded-spectrum method significantly more expensive than the other two approaches, as illustrated in Fig.~\ref{fig:performance}c.

\section{Discussion}\label{sec:discussion}

To our knowledge, this is the first work that introduces a dissipative dynamics based method for excited-state quantum chemistry. By appropriately engineering jump operators, the dynamics can be designed to drive the system toward a desired excited state, even when the initial state has zero overlap with the target. In contrast to adiabatic state preparation, dissipative approaches are not limited by artifacts arising from mean-field initialization, and can be more robust to noise. We proposed three complementary strategies for constructing such dynamics: symmetry-based, folded-spectrum, and spectral-projector approaches. Numerical benchmarks on atomic spectra, small molecules, and active-space models of benzene and ferrocene demonstrate that the framework is capable of capturing excited states with pronounced multireference and multiconfigurational character.

The effectiveness of dissipative state preparation relies on the ergodicity of the dynamics. As illustrated by the carbon atom example, restricting to a set of quadratic coupling operators can lead to non-ergodic dynamics, in which the system becomes trapped in unintended dark states that are dynamically disconnected from the target. Such obstructions can arise from symmetry constraints or rank mismatch in the coupling operator set. We show that ergodicity can be systematically restored by augmenting the operator set. For the carbon atom example, the quartic operators reintroduce the two-body scattering processes required to connect otherwise disjoint symmetry sectors. For many quantum chemistry applications, we expect that similar systematic constructions can ensure ergodicity without requiring an excessively large or unstructured set of coupling operators. An alternative and complementary strategy is to introduce controlled thermal excitation channels, which can facilitate escape from unfavorable dynamical basins.

Looking forward, the preparation of excited states in larger and more complex systems, such as condensed matter models, may present additional challenges. We keep in mind that, in order to ensure efficient simulation, the coupling operators should be local, and consequently the jump operators constructed from them should also be quasi-local. However, in these settings, low-energy competing states may differ from the target state by more than a small number of quasi-local excitations,  making strict ergodicity toward a single target state difficult to achieve. In such cases, a hybrid strategy may be more effective. Rather than enforcing convergence to a single eigenstate, dissipative dynamics can serve as a robust cooling mechanism to confine the system to a low-energy manifold. Once restricted to this manifold, quantum-phase-estimation-type techniques may be applied to resolve and prepare specific eigenstates. Moreover, recent works~\cite{DingLinYangZhang2025, LiOngLinEtAl2025} show that when access to a suitable low-energy subspace is available, phase-estimation-based methods can be generalized to exploit this subspace structure directly, and the cost can be much smaller than the inverse eigenvalue gap. Together, these directions suggest promising avenues for extending dissipative approaches to excited-state preparation and related quantum simulation tasks.

\section*{Acknowledgements}
This material is based upon work supported by the U.S. Department of Energy, Office of Science, Accelerated Research in Quantum Computing Centers, Quantum Utility through Advanced Computational Quantum Algorithms, grant no. DE-SC0025572 (H.L., L.L.), and by the Challenge Institute for Quantum Computation (CIQC) funded by National Science Foundation (NSF) through grant number OMA-2016245 (L.L.). L.L. is a Simons Investigator in Mathematics. The authors thank Garnet Chan, Gil Goldshlager, Zhen Huang, Jerome Lloyd, Avijit Shee and Yilun Yang for helpful discussions.

\section*{Supporting Information}
Additional resource estimates for the folded-spectrum and spectral-projector protocols; Monte Carlo trajectory simulation details; spin-operator expressions; Hartree--Fock degeneracy and Coulson--Fischer analyses relevant to adiabatic state preparation; depolarizing-noise robustness tests; active-space construction details for benzene and ferrocene; additional discussion of the carbon atom and augmented coupling operators; detailed numerical results for atomic and molecular excited states; and molecular geometries. This information is available free of charge via the Internet at \url{http://pubs.acs.org}.

\section*{Data and Software Availability}
The data and codes that support this study are available on GitHub via \url{https://github.com/haoen2021/LindbladAbInitio}.

\printbibliography

\newpage

\rule{0.05in}{1.75in}%
\begin{minipage}[b][1.75in]{3.25in}
  \sffamily
  \frenchspacing
  \centering
  \includegraphics[width=\linewidth]{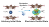}
\end{minipage}%
\rule{0.05in}{1.75in}

\clearpage


\section*{Supplementary material}

## Supplementary materials

Supplementary Text

Figures S1 to S6

Tables S1 to S7

# Supplementary Materials for

# Dissipative quantum algorithms for excited-state quantum chemistry

Hao-En Li[1], Lin Lin[1,2]*

[1]Department of Mathematics, University of California, Berkeley, California 94720, USA

[2]Applied Mathematics and Computational Research Division, Lawrence Berkeley National Laboratory, Berkeley, California 94720, USA

*Corresponding author. Email: linlin@math.berkeley.edu

**This PDF file includes:**

Supplementary Text

Figures S1 to S6

Tables S1 to S7

# S1 Additional results on hydrogen molecules

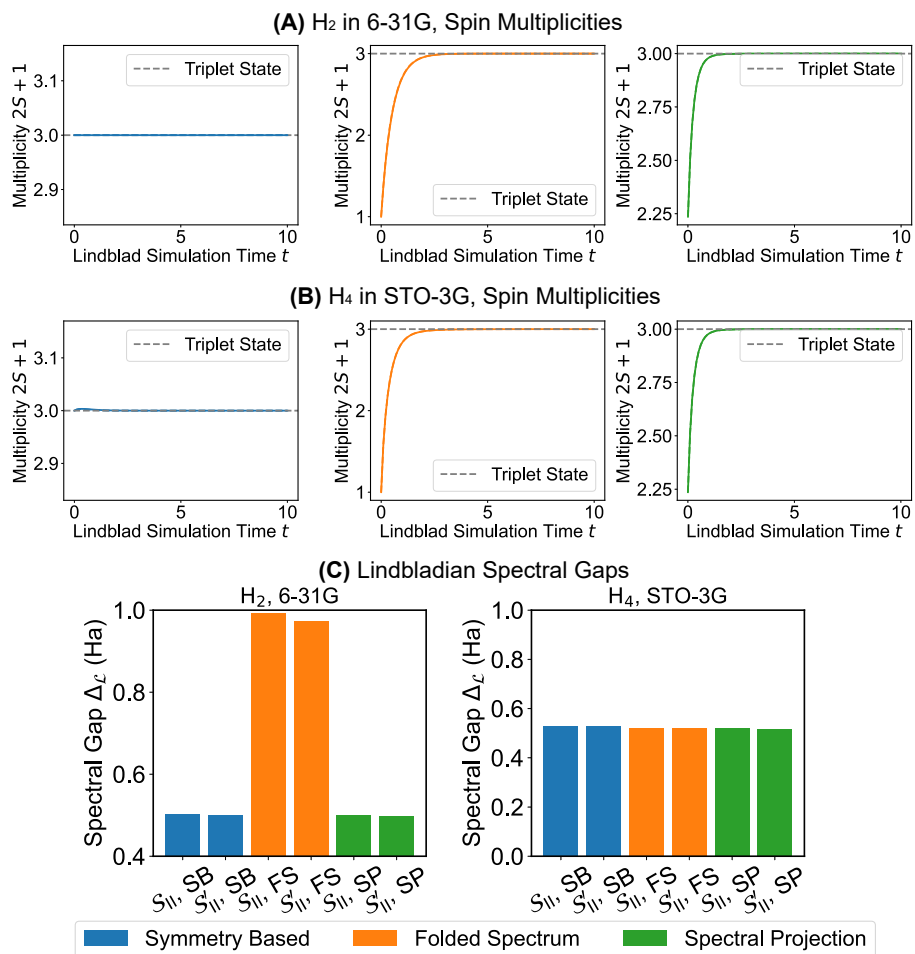

**Figure S1**: **Further comparison of the three protocols for preparing the $T_0$ and $T_0^{\pm}$ states for $H_2$ in 6-31G and $H_4$ in STO-3G.** (**A**) The convergence of the spin multiplicity for preparing the $T_0$ and $T_0^{\pm}$ states of $H_2$ (6-31G) using the three different protocols. (**B**) The convergence of the spin multiplicity for preparing the $T_0$ and $T_0^{\pm}$ states of $H_4$ (STO-3G) using the three different protocols. (**C**) The Lindbladian gap for these two systems

In this section, we provide additional results on resource estimate for comparing the total Hamiltonian simulation cost of different excited-state preparation protocols within the dissipative state preparation framework.

In Fig. S1a and b, we plot the convergence of the spin multiplicity for preparing the $T_0$ and $T_0^{\pm}$ states of $H_2$ (6-31G) and $H_4$ (STO-3G) using the three different protocols. We observe that all three

methods can successfully reproduce the correct spin multiplicities of the target states in the first excited manifold.

The spectral gap of the Lindbladian $\mathcal{L}$ is defined as

$$\Delta_{\mathcal{L}} = -\max_{\lambda \in \text{Spec}(\mathcal{L}), \lambda \neq 0} \text{Re}(\lambda). \tag{S1}$$

Briefly, the inverse spectral gap of the Lindblad generator provides an upper bound on the mixing time $T$, up to logarithmic factors in the error tolerance and constants determined by the stationary state (27, 70, 71). We plot the Lindbladian gaps for the three different excited-state preparation protocols in Fig. S1c. Here, for the symmetry-based methods, the Lindbladian gap is computed within the symmetry sector. In this case, $\Delta_{\mathcal{L}} = \Delta_{\mathcal{L}|_{\mathcal{F}_{(N_\alpha, N_\beta)}}}$, where $\mathcal{F}_{(N_\alpha, N_\beta)}$ is the configuration space sector with the correct $(N_\alpha, N_\beta)$.

Although the spectral gap of the folded-spectrum Hamiltonian is always less than or equal to that of the effective Hamiltonian within a given symmetry sector or above the reference energy $\mu$, the folded-spectrum method permits transitions from *all* states to the target excited state. In contrast, the symmetry-based and projection-based methods restrict transitions only to states with higher energies than the target excited state. As a result, the folded-spectrum method can achieve faster convergence with respect to the Lindblad simulation time. For example, in Fig. 2a in the main text, both the folded-spectrum and projection-based methods start the Lindblad dynamics from the Hartree–Fock ground state, which has a significant overlap with the many-body ground state. In the folded-spectrum method, transitions from the ground state to the target excited state are allowed, leading to a rapid decrease in infidelity over a short simulation time. In contrast, in the symmetry-based and projection-based methods, the ground state is either a dark state or projected out, so only transitions from higher-energy states to the target excited state are permitted. We also find that for $H_2$, the Lindbladian simulation time required to reach chemical accuracy with the folded-spectrum method is shorter than for the symmetry-based and spectral-projection methods, and the corresponding Lindbladian gap is also larger, consistent with the observed convergence behavior.

## S2 Classical simulation of Lindblad dynamics via Monte Carlo trajectory method

The direct integration of the Lindblad equation requires propagating the many-body density operator, which has a size of $\dim(\mathcal{F}) \times \dim(\mathcal{F})$ where $\mathcal{F}$ is the configuration space of the system, or the sector of the Fock space with correct spin and particle number. This can be already computationally prohibitive for system sizes beyond a few spin orbitals. An alternative approach is to simulate the unraveled Lindblad dynamics using a Monte Carlo trajectory method (*72*). In this approach, the many-body density operator $\rho(t)$ is approximated by an ensemble average of $N_{\text{traj}}$ pure state trajectories $\psi_t$. Therefore, instead of propagating the density operator, we only need to propagate $N_{\text{traj}}$ wavefunctions, each of size $\dim(\mathcal{F})$, which can significantly reduce the computational cost. We refer readers to (*72, 73*) for more details about the Monte Carlo trajectory method for simulating Lindblad dynamics. In this work, we employ the quantum jump method, where each wavefunction trajectory $\psi(t)$ evolves according to a non-Hermitian effective Hamiltonian

$$H_{\text{eff}} = H - \frac{i}{2} \sum_k K_k^\dagger K_k, \tag{S2}$$

interrupted by stochastic quantum jumps induced by the jump operators $\{K_k\}$. Since $\sum_k K_k^\dagger K_k$ is positive semidefinite, the non-Hermitian term $-\frac{i}{2} \sum_k K_k^\dagger K_k$ leads to a decay in the norm of the wavefunction

$$\left\| e^{-iH_{\text{eff}}\Delta t} \psi_t \right\|^2 = \|\psi_t\|^2 - \Delta t \sum_k \|K_k \psi_t\|^2 + O(\Delta t^2) \tag{S3}$$

over a small time interval $\Delta t$. This decay can be interpreted as, the state has the probability $\|\psi_t\|^2 - \Delta t \sum_k \|K_k \psi_t\|^2$ (which is in $[0, 1]$ for $\Delta t$ small) to remain unchanged. Otherwise, the wavefunction undergoes a quantum jump. The probability of a quantum jump occurring due to the $k$-th jump operator is given by

$$p_k = \frac{\Delta t \|K_k \psi_t\|^2}{\Delta t \sum_k \|K_k \psi_t\|^2} = \frac{\|K_k \psi_t\|^2}{\sum_k \|K_k \psi_t\|^2}. \tag{S4}$$

And if this occurs, the wavefunction undergoes a "quantum jump" into the state defined by the action of the jump operator:

$$\psi_{t+\Delta t} = \frac{K_k \psi_t}{\|K_k \psi_t\|} \tag{S5}$$

after which the wavefunction is renormalized to have unit norm and the evolution continuous. This can be understood as the numerical solution to the following Poisson-type stochastic differential equation

$$d\psi_t = -iH_{\text{eff}}\psi_t dt + \sum_k \left(\frac{K_k\psi_t}{\|K_k\psi_t\|} - \psi_t\right) dN_k(t) \tag{S6}$$

where $N_k(t)$ are independent Poisson processes with intensities $\lambda_k(t) = \|K_k\psi_t\|^2$. By the Itô calculus for the Poisson processes, it can be verified that $\rho(t) := \mathbb{E}[|\psi_t\rangle\langle\psi_t|]$ solves the original Lindblad equation. Thus numerically, we can sample $N_{\text{traj}}$ independent trajectories $\psi_t^{(i)}$ and approximate the density operator by

$$\rho(t) \approx \frac{1}{N_{\text{traj}}} \sum_{i=1}^{N_{\text{traj}}} \psi_t^{(i)} \psi_t^{(i)\dagger}. \tag{S7}$$

For each trajectory, we can use the following algorithm to propagate the wavefunction $\psi_t$. We first choose a random number $R_1$ uniformly from $[0, 1]$ to determine the probability that a quantum jump occurs. We then choose a random number $R_2$ uniformly from $[0, 1]$ to determine which jump operator causes the quantum jump if it occurs. We then integrate the effective Schrödinger equation according to $H_{\text{eff}}$ with time step $\Delta$, until a time $\tau$ such that $\|\psi_\tau\|^2 \leq R_1$, at which point a quantum jump occurs. Then the wavefunction jumps to one of the normalized state given in Eq. S5, with probability given in Eq. S4, according to $R_2$. After the jump, we renormalize the wavefunction and continue the evolution until the final simulation time is reached.

In our state-preparation dissipative dynamics, quantum jumps tend to occur infrequently. As a result, many sampled trajectories correspond to identical non-jumping deterministic evolutions. To reduce this redundancy, we apply the strategy from (*74*), which allows the non-jumping trajectory to be sampled only once. The non-jumping probability $p$ is first determined, and for subsequent runs of trajectories, we only generate random numbers $R_1$ constrained by $R_1 > p$ to guarantee that a jump takes place. When computing observables, we assign a weight $p$ to the non-jumping trajectory and $1 - p$ to those involving jumps. In QuTiP, this method is available through the "improved_sampling" option (*75*).

## S3 Expression of spin operators

In this section, we provide the expressions of the spin observables. Consider the system with $L$ spatial molecular orbitals, the number of $\alpha$ and $\beta$ electrons are denoted as $N_\alpha$ and $N_\beta$ respectively. The $(N_\alpha, N_\beta)$-sector of the Fock space is invariant under the action of the spin square operator $\hat{S}^2$.

The spin square operator is defined as

$$\hat{S}^2 = \hat{S}_z^2 + \hat{S}_+\hat{S}_- + \hat{S}_-\hat{S}_+, \tag{S8}$$

where

$$\hat{S}_+ = \frac{1}{\sqrt{2}}(\hat{S}_x + i\hat{S}_y), \quad \hat{S}_- = \frac{1}{\sqrt{2}}(\hat{S}_x - i\hat{S}_y) \tag{S9}$$

and we use the spinor expression of the spin operators

$$\hat{S}_x = \frac{1}{2}\begin{pmatrix}\mathbf{a}_\alpha^\dagger & \mathbf{a}_\beta^\dagger\end{pmatrix}\sigma_x\begin{pmatrix}\mathbf{a}_\alpha \\ \mathbf{a}_\beta\end{pmatrix}, \quad \hat{S}_y = \frac{1}{2}\begin{pmatrix}\mathbf{a}_\alpha^\dagger & \mathbf{a}_\beta^\dagger\end{pmatrix}\sigma_y\begin{pmatrix}\mathbf{a}_\alpha \\ \mathbf{a}_\beta\end{pmatrix}, \quad \hat{S}_z = \frac{1}{2}\begin{pmatrix}\mathbf{a}_\alpha^\dagger & \mathbf{a}_\beta^\dagger\end{pmatrix}\sigma_z\begin{pmatrix}\mathbf{a}_\alpha \\ \mathbf{a}_\beta\end{pmatrix}, \tag{S10}$$

with

$$\sigma_x = \begin{pmatrix}0 & I \\ I & 0\end{pmatrix}, \quad \sigma_y = \begin{pmatrix}0 & -iI \\ iI & 0\end{pmatrix}, \quad \sigma_z = \begin{pmatrix}I & 0 \\ 0 & -I\end{pmatrix}, \tag{S11}$$

and

$$\mathbf{a}_\alpha^\dagger = \begin{pmatrix}a_{1\alpha}^\dagger & a_{2\alpha}^\dagger & \cdots & a_{L\alpha}^\dagger\end{pmatrix}, \quad \mathbf{a}_\beta^\dagger = \begin{pmatrix}a_{1\beta}^\dagger & a_{2\beta}^\dagger & \cdots & a_{L\beta}^\dagger\end{pmatrix}, \tag{S12}$$

which are the fermionic creation operators in the atomic orbital (AO) basis, which is assumed to be orthonormal for simplicity. In our study, we consider the spin-attributed wavefunction i.e. the wavefunction is restricted in the sectors with fixed $N_\alpha - N_\beta$ (which is the `spin` parameter in `PySCF`).

After performing the Hartree–Fock calculation, we can express the spin operators in terms of the fermionic creation and annihilation operators in the MO basis via a basis transformation of the AO operators. The fermionic creation operators in the MO basis are given by

$$\mathbf{c}_\alpha^\dagger = (c_{1\alpha}^\dagger, c_{2\alpha}^\dagger, \cdots, c_{L\alpha}^\dagger) = \mathbf{a}_\alpha^\dagger \Phi_\alpha, \quad \mathbf{c}_\beta^\dagger = (c_{1\beta}^\dagger, c_{2\beta}^\dagger, \cdots, c_{L\beta}^\dagger) = \mathbf{a}_\beta^\dagger \Phi_\beta, \tag{S13}$$

where the unitary matrices $\Phi_\alpha$ and $\Phi_\beta$ are the MO coefficient matrices for $\alpha$ and $\beta$ spins respectively. We denote the overlap matrices between the $\alpha$ and $\beta$ MO coefficient matrices as

$$M = \Phi_\alpha^\dagger \Phi_\beta, \quad M^\dagger = \Phi_\beta^\dagger \Phi_\alpha. \tag{S14}$$

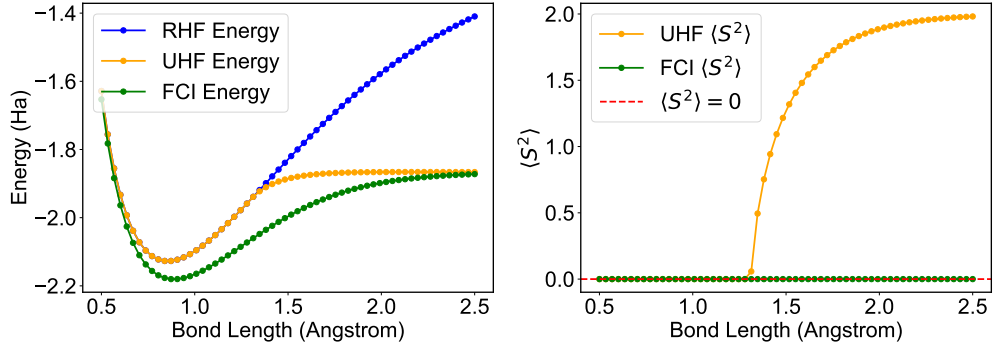

**Figure S2**: **Comparison between RHF and UHF methods for the $H_4$ system along the bond dissociation pathway.** The solid lines represent the potential energy curves obtained from the RHF (blue) and UHF (orange) methods. The Coulson–Fischer point is observed at a bond length of approximately 1.31 Å, where the UHF solution diverges from the RHF solution, providing a more accurate description of the energy as it approaches dissociation. We also see that after the Coulson–Fischer point, the expectation value of the total spin operator $\hat{S}^2$ increases for the UHF solution, indicating a higher degree of spin contamination.

Using the above notations, we can express the spin square operator as

$$\hat{S}^2 = -\sum_{i,j,k,l=1}^{L} M_{ij}(M^\dagger)_{kl} c_{i\alpha}^\dagger c_{l\alpha} c_{k\beta}^\dagger c_{j\beta} + \frac{N_\alpha + N_\beta}{2} + \frac{(N_\alpha - N_\beta)^2}{4}. \tag{S15}$$

In particular, for the RHF case where $M = I$, we have a simplified expression

$$\hat{S}^2 = -\sum_{i,j=1}^{L} c_{i\alpha}^\dagger c_{j\alpha} c_{j\beta}^\dagger c_{i\beta} + \frac{N_\alpha + N_\beta}{2} + \frac{(N_\alpha - N_\beta)^2}{4}. \tag{S16}$$

## S4 Review of Hartree–Fock methods and the degeneracy issue

The Hartree–Fock theory aims to find a self-consistent single-particle operator approximation to the full many-body Hamiltonian, taking the form (in the second quantization representation)

$$H_{\text{HF}} = \sum_{p,q=1}^{L} \sum_{\sigma=\alpha,\beta} F_{pq,\sigma} a_{p\sigma}^\dagger a_{q\sigma}, \tag{S17}$$

where $a_{p\sigma}^\dagger$ and $a_{q\sigma}$ are the fermionic creation and annihilation operators for the *atomic* spatial orbital $p, q$ with spin $\sigma$. We assume that the atomic orbitals are orthogonalized. The Fock matrix

$F = F_\alpha \oplus F_\beta$ is given by

$$F_{\sigma,pq}[D] = h_{pq} + V_{\sigma,pq}[D] - K_{\sigma,pq}[D], \tag{S18}$$

where $h_{pq}$ is a fixed single-particle matrix and is independent of the spin, so we omit the spin index hereafter. The direct Coulomb term $V$ and the Fock exchange term $K$ depend nonlinearly on the 1-RDM $D$, which is a $2L \times 2L$ matrix in the spin-orbital basis with a block-diagonal structure:

$$D = D_\alpha \oplus D_\beta, \quad (D_\sigma)_{pq} = \sum_{i=1}^{N_\sigma} \Phi_{\sigma,pi} \Phi^*_{\sigma,qi}, \quad \sigma \in \{\alpha, \beta\} \tag{S19}$$

Here $\Phi = \Phi_\alpha \oplus \Phi_\beta$ and $\Phi_\sigma \in \mathbb{C}^{L \times L}$ is a unitary matrix called the molecular orbital coefficient matrix for spin $\sigma$, which are eigenvectors of the Fock matrix $F_\sigma$. Therefore this yields a pair of coupled self-consistent eigenvalue problems:

$$F_\sigma[D]\Phi_\sigma = \Phi_\sigma \varepsilon_\sigma, \quad \sigma \in \{\alpha, \beta\}. \tag{S20}$$

After the self-consistency is achieved, we obtain the linearized Hartree–Fock Hamiltonian Eq. S17. This Hamiltonian is quadratic in the fermionic operators and its ground state has an explicit form as a Slater determinant:

$$|\text{HF}\rangle = \prod_{i=1}^{N_\alpha} c^\dagger_{i\alpha} \prod_{j=1}^{N_\beta} c^\dagger_{j\beta} |0\rangle, \tag{S21}$$

where $|0\rangle$ is the physical vacuum and the new set of fermionic operators $\{c^\dagger_{p\sigma}, c_{p\sigma}\}$ are defined through the molecular orbital basis transformation:

$$c^\dagger_{p\sigma} = \sum_{q=1}^{L} \Phi_{\sigma,qp} a^\dagger_{q\sigma}, \quad c_{p\sigma} = \sum_{q=1}^{L} \Phi^*_{\sigma,qp} a_{q\sigma}. \tag{S22}$$

In the restricted Hartree–Fock (RHF) method, the density matrices for the two spin channels are constrained to be identical. Consequently, only a single self-consistent eigenvalue problem is solved. This produces a spin-independent Fock matrix. In contrast, the unrestricted Hartree–Fock (UHF) method allows the two spin channels to have distinct density matrices, which leads to a pair of coupled self-consistent eigenvalue problems and, in general, a block-diagonal Fock matrix of the form $F = F_\alpha \oplus F_\beta$.

For closed-shell molecules, however, the UHF solution often collapses to the restricted limit, with $F_\alpha = F_\beta$. This restoration of spin symmetry enforces degeneracies in the orbital spectrum, which in turn lead to degenerate many-body excited states of the Hartree–Fock Hamiltonian $H_{\mathrm{HF}}$. Such degeneracies pose a direct challenge for adiabatic state preparation, since they can induce gap closures along the adiabatic path and cause the evolution to converge to an uncontrolled superposition of configurations. We illustrate this issue using the $H_4$ system in its near-equilibrium geometry ($R = 0.7$ Å). As shown in Fig. S3, even when starting from a UHF formalism, the initial Hamiltonian exhibits degenerate levels at the beginning of the adiabatic evolution.

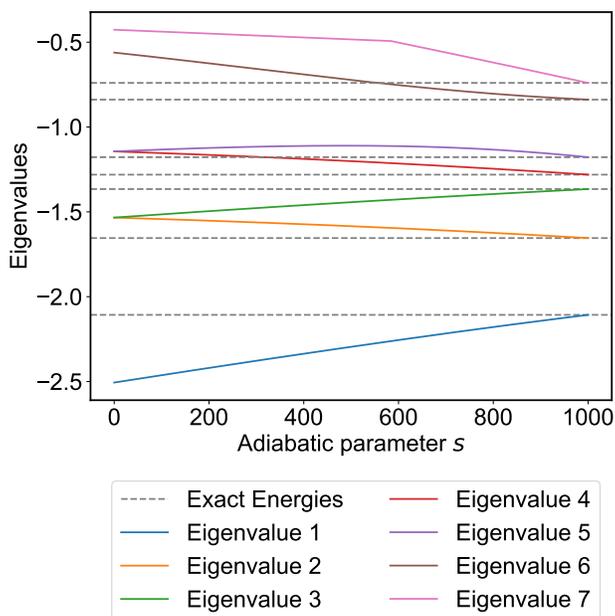

Figure S3: **The eigenvalue spectrum of the adiabatic path Hamiltonian for $H_4$ at bond length 0.7 Å.** The initial Hamiltonian is chosen to be the UHF Hamiltonian and the final Hamiltonian is the FCI Hamiltonian in the same molecular orbital basis. We can see the levels are crossed at the beginning of the adiabatic path Eq. S24.

S9

# S5 Coulson–Fischer point and robustness of dissipative state preparation along bond dissociation

In hydrogen chain systems, the Coulson–Fischer (CF) point marks the bond length at which the Restricted Hartree–Fock (RHF) solution becomes unstable, transitioning from a local minimum to a saddle point. Beyond this point, an Unrestricted Hartree–Fock (UHF) solution of lower energy emerges. This instability arises because the RHF framework constrains the spatial orbitals of $\alpha$ and $\beta$ electrons to be identical. This is a valid approximation near equilibrium geometries, but one that fails to capture the static correlation intrinsic to bond breaking.

The UHF method lowers the total energy by relaxing this constraint, allowing spatial orbitals to differ and thereby breaking spin symmetry. While this provides a more accurate potential energy surface during dissociation, the resulting UHF wavefunction is no longer an eigenstate of the total spin operator $\hat{S}^2$. This phenomenon, known as spin contamination, implies the solution is physically mixed; as illustrated in Fig. S2, the expectation value $\langle \hat{S}^2 \rangle$ for the UHF solution deviates significantly from the exact singlet value (0) beyond the CF point. Although the utility of symmetry breaking in electronic structure remains a subject of debate (*76*), we focus here on comparing the robustness of Dissipative State Preparation (DSP) against Adiabatic State Preparation (ASP) across this challenging regime.

To access the symmetry-broken UHF solution within the PySCF framework, we must explicitly break the spin symmetry of the initial guess, as the default delocalized density matrix typically converges to the unstable RHF solution. For example, for $H_4$ (STO-3G), we induce spin polarization by initializing the density matrix ($D$) with localized electron assignments:

$$D = \begin{pmatrix} 1 & 0 & 0 & 0 \\ 0 & 1 & 0 & 0 \\ 0 & 0 & 0 & 0 \\ 0 & 0 & 0 & 0 \end{pmatrix}_\alpha \oplus \begin{pmatrix} 0 & 0 & 0 & 0 \\ 0 & 0 & 0 & 0 \\ 0 & 0 & 1 & 0 \\ 0 & 0 & 0 & 1 \end{pmatrix}_\beta. \quad \text{(S23)}$$

This initialization ensures convergence to the lower-energy UHF solution that properly reproduces the Coulson–Fischer point (see Fig. S2).

Using this converged UHF Hamiltonian ($H_{\text{UHF}}$) and its corresponding ground state $\Phi_{\text{UHF}} = \Phi_{\text{UHF},\alpha} \oplus \Phi_{\text{UHF},\beta}$, we construct the molecular orbital basis. We then define the adiabatic path

Hamiltonian for the ASP method as:

$$H(s/T) = \left(1 - \frac{s}{T}\right) H_{\text{UHF}} + \frac{s}{T} H_{\text{FCI}}, \qquad (S24)$$

where $H_{\text{FCI}}$ is the full configuration interaction Hamiltonian in the same basis. The evolution initializes in the UHF ground state. For comparison, the dissipative state preparation (DSP) method is also initialized from the same UHF ground state.

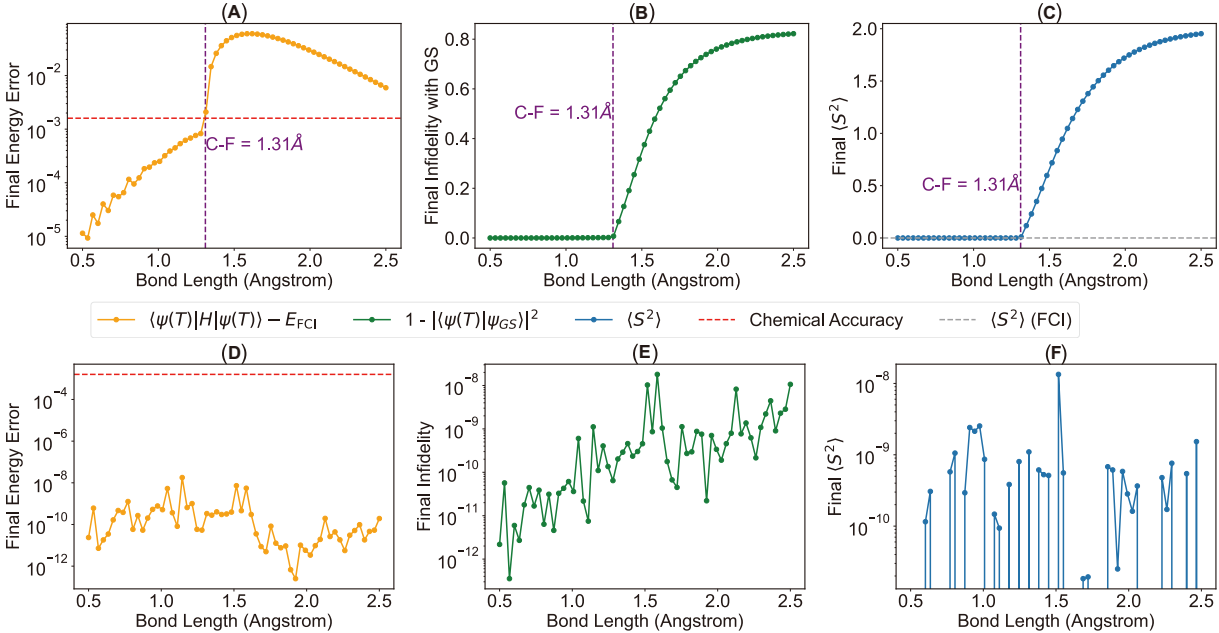

**Figure S4**: **Comparison of Dissipative State Preparation (DSP) and Adiabatic State Preparation (ASP) for H$_2$ (6-31G) and H$_4$ (STO-3G).** (**A**) Final energy errors for ASP. The vertical dashed purple line indicates the Coulson–Fischer point. Beyond this transition ($R \approx 1.31$ Å), ASP fails to achieve chemical accuracy. (**B**) State infidelity ($1 - F$) for ASP, showing an abrupt increase in error after the critical bond length. (**C**) Expectation value $\langle \hat{S}^2 \rangle$ for the ASP-prepared state. The significant deviation from 0 indicates that ASP preserves the spin contamination of the mean-field reference. (**D**-**F**) Performance metrics for DSP: (**D**) Energy errors, (**E**) Infidelity, and (**F**) Spin contamination. In contrast to ASP, DSP consistently maintains high accuracy, low infidelity, and correct spin properties across the entire potential energy curve.

Here, we compare the performance of the dissipative state preparation and the ASP methods for describing the potential energy curves of the equidistant H$_4$ chain in STO-3G. In both method,

we choose the stopping time $T = 30$ and the number of steps $N = 1000$. In Fig. S4, ASP exhibits a sharp degradation in performance near the CF point ($\sim 1.31$ Å): the energy error abruptly rises above the chemical-accuracy threshold; the infidelity increases rapidly; and the final state shows pronounced spin contamination, deviating from the singlet character of the true FCI ground state. These signatures indicate that once the mean-field reference becomes unstable and breaks symmetry, ASP fails to reliably track the correct correlated ground state. In contrast, the dissipative state preparation (DSP) method remains robust across the entire dissociation curve, achieving energy errors well within chemical accuracy, maintaining high fidelity, and accurately recovering the singlet spin properties with $\langle \hat{S}^2 \rangle \approx 0$ as shown in Fig. S4d–f.

## S6  Robustness against depolarizing noise

In this section, we investigate the robustness of the DSP against depolarizing noise and compare it with the ASP method. The depolarizing semigroup in the Schrödinger picture is given by

$$\mathcal{D}_t^\gamma[\rho] = e^{-\gamma t}\rho + (1 - e^{-\gamma t})\frac{I}{d}, \tag{S25}$$

which is a self-adjoint superoperator with respect to the Hilbert–Schmidt inner product. $\gamma > 0$ is the depolarizing rate. $d$ is the dimension of the underlying Hilbert space. The corresponding generator is

$$\mathcal{L}_{\text{dep}}^\gamma[\rho] = \gamma \left( \text{Tr}(\rho)\frac{I}{d} - \rho \right). \tag{S26}$$

Note that this generator can also be written in the Lindblad form, and the expression is not unique.

We investigate two different quantum semigroup generators for the adiabatic state preparation and dissipative state preparation, with the depolarizing noise added:

$$\mathcal{L}_{\text{ASP}}(s)[\rho] = -\text{i}[H(s/T), \rho] + \mathcal{L}_{\text{dep}}^\gamma[\rho], \tag{S27}$$

where $H(s/T)$ is the adiabatic path Hamiltonian,

$$H\left(\frac{s}{T}\right) = \left(1 - \frac{s}{T}\right) H_{\text{UHF}} + \frac{s}{T} H_{\text{FCI}}, \tag{S28}$$

and correspondingly for the dissipative state preparation, the generator with depolarizing noise is

$$\mathcal{L}_{\text{DSP}}[\rho] = -\text{i}[H, \rho] + \sum_k \left( K_k \rho K_k^\dagger - \frac{1}{2}\{K_k K_k^\dagger, \rho\} \right) + \mathcal{L}_{\text{dep}}^\gamma[\rho]. \tag{S29}$$

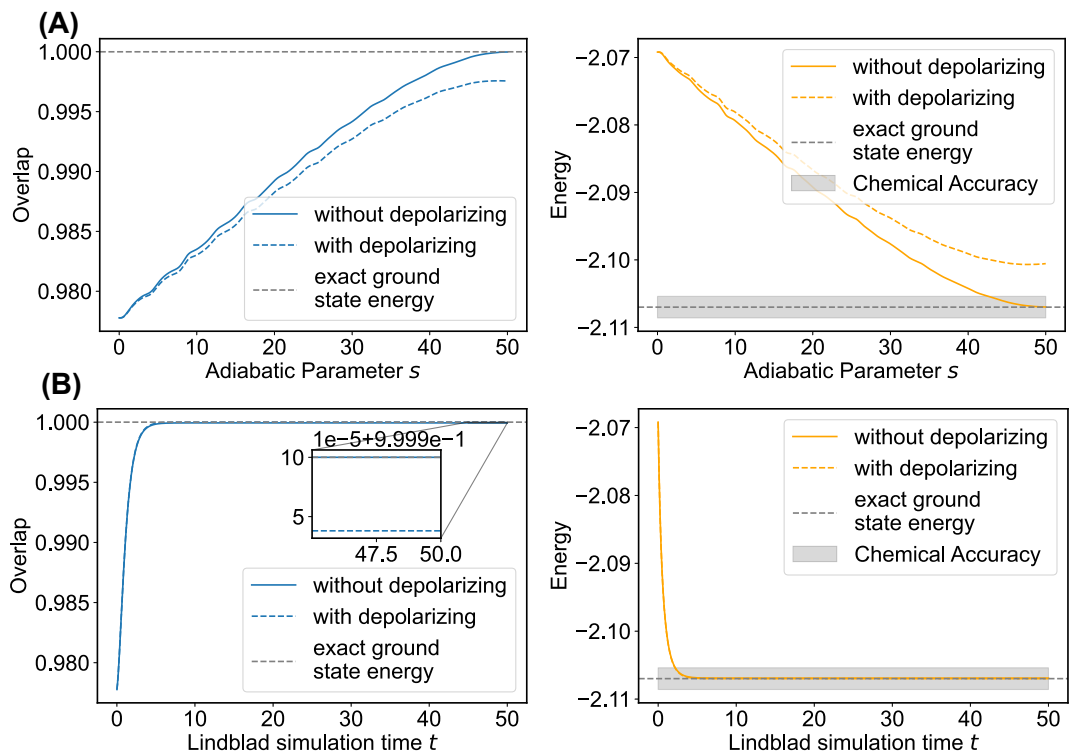

**Figure S5**: **Robustness of dissipative state preparation against depolarizing noise.**(**A**) The adiabatic state preparation from the Hartree–Fock state to the FCI ground state of the $H_4$ chain in STO-3G and the bondlength 0.7 Å. The solid lines represent the results without depolarizing noise, while the dashed lines represent the results with depolarizing noise of rate $\gamma = 5 \times 10^{-5}$. We can see that the final fidelity is significantly reduced due to the depolarizing noise and the final energy fails to reach the chemical accuracy threshold (shaded area) with the noise. (**B**) The dissipative state preparation from the Hartree–Fock state to the FCI ground state of the same system using the Type-II set of jump operators. The solid lines represent the results without depolarizing noise, while the dashed lines represent the results with depolarizing noise of rate $\gamma = 5 \times 10^{-5}$. We can see that even with the depolarizing noise, the final fidelity does not change much and the accuracy of the final energy is still far below the chemical accuracy threshold (shaded area).

We consider the same equidistant H$_4$ chain system in STO-3G at bondlength 0.7 Å as in Supplementary Note S5, and choose the UHF Hamiltonian as the initial Hamiltonian for the ASP method. The dissipative state preparation uses the Type-II set of jump operators. The stopping time is chosen to be $T = 50$. The depolarizing rate is chosen to be $\gamma = 5 \times 10^{-5}$. The numerical results are summarized in Fig. S5. We can see that for the ASP method, the final fidelity is significantly reduced due to the depolarizing noise and the final energy fails to reach the chemical accuracy threshold with the noise, as shown in Fig. S5a. In contrast, for the dissipative state preparation method, even with the depolarizing noise, the final fidelity does not change much, as shown in Fig. S5b. This demonstrates the superior robustness of the dissipative state preparation method against depolarizing noise compared to the adiabatic state preparation method.

## S7 Construction of the active space using AVAS and PiOS

For constructing the active space for ferrocene, we employ the atomic valence active space (AVAS) method (*39*), which is a well-defined procedure to constructing active spaces based on the projectors onto atomic valence orbitals. The goal of the AVAS procedure is to identify the set $\mathcal{B}$ of valence atomic orbitals with significant overlap with the Hartree–Fock occupied orbitals $|i\rangle$ and virtual orbitals $|a\rangle$. The Hartree–Fock molecular orbitals are linear combinations of some basis functions from some computational basis set, which in general do not correspond directly to any sort of atomic orbitals. To bridge this gap, the AVAS method constructs the projector onto the $\mathcal{B}$ space as

$$P_{\mathcal{B}} = \sum_{|p\rangle,|q\rangle \in \mathcal{B}} [S^{-1}]_{pq} |p\rangle \langle q|, \quad (S30)$$

where $S = (\langle p|q\rangle)_{p,q}$ is the overlap matrix of the atomic orbitals in $\mathcal{B}$. Then we construct two overlap matrices of the occupied HF orbital space and the virtual HF orbital space respectively as

$$S^{\mathcal{B}}_{\text{occ}} = (\langle i|P_{\mathcal{B}}|j\rangle)_{i,j}, \quad S^{\mathcal{B}}_{\text{virt}} = (\langle a|P_{\mathcal{B}}|b\rangle)_{a,b}. \quad (S31)$$

In the construction of $S^{\mathcal{B}}_{\text{occ}}$, we freeze the core occupied orbitals and later add them back to the inactive (environment) space. We then diagonalize these two overlap matrices to obtain their eigenvalues and eigenvectors. The eigenvectors with non-negligible eigenvalues (with respect to some threshold) correspond to the occupied and virtual orbitals that have significant overlap with

the atomic orbitals in $\mathcal{B}$. By collecting these orbitals, which are rotated from the original $|i\rangle$ and $|a\rangle$, we choose the active space for subsequent many-body calculations.

For constructing the active space for benzene, we employ the $\pi$-orbital space (PiOS) method (*40*). The key idea of PiOS method is similar to that of AVAS, but it focuses on selecting the $\pi$ orbitals, and thus only the $p_z$ orbitals perpendicular to the molecular plane instead of all the valence atomic orbitals are included in the set $\mathcal{B}$.

In practice, the set of atomic orbitals $\mathcal{B}$ is usually chosen to be the minimal tabulated free-atom AOs, such as MINAO, or the intrinsic atomic orbitals (IAO) that correspond to the chemical atomic orbitals while also account for the polarization effects from the molecular environment. In this work, we choose the MINAO basis set for the valence atomic orbitals, and choose some larger computational basis set for the Hartree–Fock calculation.

For benzene, we first perform a restricted Hartree–Fock (RHF) calculation within the aug-cc-pvTZ basis set to obtain the molecular orbitals, at the geometry previously optimized at the UB3LYP/cc-pvTZ level. Then the PiOS procedure selects six $\pi$ orbitals to define a (6e, 6o) active space. Then a CASSCF calculation with state-averaging over the lowest seven singlet states (including the ground state) is performed to optimize the active-space orbitals. The detailed technical settings can be found in Ref. (*40*).

For ferrocene, an initial restricted open-shell Hartree–Fock (ROHF) calculation was performed within the cc-pvTZ-DK basis set to obtain the molecular orbitals, at the $D_{5h}$-symmetric geometry previously optimized at the CCSD(T)/cc-pwCVTZ level (*77*), where both cyclopentadienyl rings are planar and in the eclipsed conformation, and the $z$-axis coincides with the Cp-Fe-Cp axis. Using this geometry, an AVAS procedure with a 0.1 threshold was applied to the five Fe $3d$ orbitals, selecting five occupied and two virtual orbitals to define a (10e, 7o) active space. For the technical details and the visualization of the active orbitals, we refer readers to Ref. (*39*).

For the visualization of the EDD of the excited states in Fig. 5a, we first compute the one-body reduced density matrix (1-RDM) $D$ of the converged steady-state density matrix $\rho$ of the Lindblad dynamics corresponding to the target excited state, in the MO basis:

$$D_{ij} = \text{Tr}\left(\rho c_j^\dagger c_i\right). \tag{S32}$$

Then we transform the 1-RDM to the AO basis via the MO coefficient matrix $\Phi$ generated from the

PiOS or AVAS procedure:

$$D^{AO} = \Phi D \Phi^\dagger. \tag{S33}$$

Here $\Phi \in \mathbb{C}^{N_{AO} \times N_{act}}$ contains as its columns the active-space MO coefficients in terms of the AO basis, where $N_{AO}$ is the total number of AOs and $N_{act}$ is the number of active MOs, respectively. Then we subtract the 1-RDM of the CASCI ground state $D_{GS}^{AO}$ from $D^{AO}$ to obtain the EDD:

$$\Delta D^{AO} = D^{AO} - D_{GS}^{AO}. \tag{S34}$$

Finally, we visualize the EDD in the AO basis using the cubegen utility in PySCF (*41, 42*). The EDD plot is produced with VESTA.

# S8 Details about the $^1S$ and $^5S$ states of the carbon atom

We consider the carbon atom example discussed in the main text.

Note that for any $A_k \in \mathcal{S}_{II}$, we have $\langle \psi_{5S} | A_k | \psi_{1S} \rangle \approx 0$. Consequently,

$$K_k |\psi_{1S}\rangle \approx 0, \quad \forall K_k \text{ constructed from } \mathcal{S}_{II} \text{ (and therefore the subset } \mathcal{S}_{II}'\text{)}, \tag{S35}$$

because

$$K_k |\psi_{1S}\rangle = \sum_j \hat{f}\left[(\lambda_j - \mu)^2 - (\lambda_{1S} - \mu)^2\right] \langle \psi_j | A_k | \psi_{1S} \rangle |\psi_j\rangle. \tag{S36}$$

If $|\psi_j\rangle \neq |\psi_{5S}\rangle$, note that the $^1S$ state is the lowest-energy excited state lying above the $^5S$ ground state. Hence, by the construction of the filter function,

$$(\lambda_j - \mu)^2 - (\lambda_{1S} - \mu)^2 \geq \Delta^2, \tag{S37}$$

which implies $\hat{f}\left[(\lambda_j - \mu)^2 - (\lambda_{1S} - \mu)^2\right] = 0$.

If $|\psi_j\rangle = |\psi_{5S}\rangle$, then $\langle \psi_{5S} | A_k | \psi_{1S} \rangle \approx 0$ according to our numerical verification. Therefore, $K_k |\psi_{1S}\rangle \approx 0$ for all $K_k$ constructed from $\mathcal{S}_{II}$.

This indicates that the $^1S$ state is approximately another steady state of the dissipative dynamics governed by jump operators from $\mathcal{S}_{II}$. As a result, the population of the $^1S$ state remains non-decreasing during the dissipative evolution, which explains the failure to prepare the $^5S$ ground state—even when starting from an initial state with large overlap with $^5S$ and negligible overlap with $^1S$.

The quartic operators in $Q$ can effectively couple the $^1S$ and $^5S$ states, as demonstrated by the non-negligible matrix elements in Fig. S6.

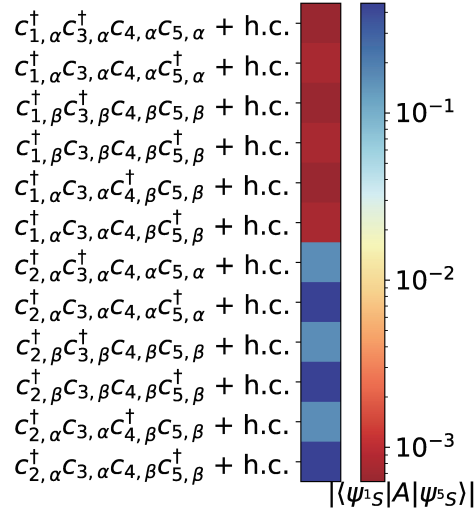

**Figure S6**: **Heatmap of the matrix elements $\left|\langle\psi_{1_S}|A|\psi_{5_S}\rangle\right|$ for various quartic operators $A$.**

# S9 Choice of the high-spin initial state and augmented set of coupling operators

We discuss the choice of the high-spin initial state for preparing the quintet excited state in the carbon atom example. The high-spin initial state $|D\rangle$ is chosen as

$$|D\rangle = \frac{1}{2}\left(\left|\Psi_{2\beta3\alpha}^{4\beta5\alpha}\right\rangle + \left|\Psi_{2\alpha3\beta}^{4\alpha5\beta}\right\rangle + \left|\Psi_{2\beta3\beta}^{4\beta5\beta}\right\rangle + \left|\Psi_{2\alpha3\alpha}^{4\alpha5\beta}\right\rangle\right). \tag{S38}$$

Here the index $i, j, a, b \in \{1, 2, 3, 4, 5\}$ are the spatial MO indices and $\sigma, \tau, \sigma', \tau' \in \{\alpha, \beta\}$ are the spin indices. We recall that the STO-3G basis set for the second-period element contains 5 spatial orbitals. The double-excited Slater determinant

$$\left|\Psi_{i\sigma j\tau}^{a\sigma' b\tau'}\right\rangle = c_{a\sigma'}^\dagger c_{b\tau'}^\dagger c_{i\sigma} c_{j\tau} |\text{HF}\rangle, \tag{S39}$$

denotes the Slater determinant obtained from removing the electrons from the occupied orbitals $i\sigma$ and $j\tau$ and placing them in the virtual orbitals $a\sigma'$ and $b\tau'$, from the HF ground state

$$|\text{HF}\rangle = c_{1\alpha}^\dagger c_{1\beta}^\dagger c_{2\alpha}^\dagger c_{2\beta}^\dagger c_{3\alpha}^\dagger c_{3\beta}^\dagger |0\rangle, \tag{S40}$$

S17

where $|0\rangle$ is the physical vacuum. The state $|D\rangle$ is a linear combination of four determinants with high spin multiplicity and has a significant overlap with the $^5S$ state ($|\langle\psi_{5S}|D\rangle|^2 > \frac{1}{2}$).

For the augmented set of coupling operators, as shown in Fig. S6, many quartic operators have significant matrix elements $\langle\psi_{1S}|A|\psi_{5S}\rangle$ between $^1S$ and $^5S$. The following set of quartic operators $Q$ is chosen to effectively couple the $^1S$ and $^5S$ states in the carbon atom example:

$$\begin{aligned}Q = \{&c_{2\alpha}^\dagger c_{3\alpha}^\dagger c_{4\alpha} c_{5\alpha} + \text{h.c.}, \quad c_{2\alpha}^\dagger c_{3\alpha} c_{4\alpha} c_{5\alpha}^\dagger + \text{h.c.}, \quad c_{2\beta}^\dagger c_{3\beta}^\dagger c_{4\beta} c_{5\beta} + \text{h.c.},\\ &c_{2\beta}^\dagger c_{3\beta} c_{4\beta} c_{5\beta}^\dagger + \text{h.c.}, \quad c_{2\alpha}^\dagger c_{3\alpha} c_{4\beta}^\dagger c_{5\beta} + \text{h.c.}, \quad c_{2\alpha}^\dagger c_{3\alpha} c_{4\beta} c_{5\beta}^\dagger + \text{h.c.}\}.\end{aligned} \tag{S41}$$

For $H_2O$, $C_6H_6$ and $FeCp_2$ examples, we choose similar set of quartic operators that can effectively couple the high-spin eigenstates with other eigenstates as follows:

$$\begin{aligned}Q_{H_2O} = \{&c_{4\alpha}^\dagger c_{5\alpha}^\dagger c_{6\alpha} c_{7\alpha} + \text{h.c.}, \quad c_{4\alpha}^\dagger c_{5\alpha} c_{6\alpha} c_{7\alpha}^\dagger + \text{h.c.}, \quad c_{4\beta}^\dagger c_{5\beta}^\dagger c_{6\beta} c_{7\beta} + \text{h.c.},\\ &c_{4\beta}^\dagger c_{5\beta} c_{6\beta} c_{7\beta}^\dagger + \text{h.c.}, \quad c_{4\alpha}^\dagger c_{5\alpha} c_{6\beta}^\dagger c_{7\beta} + \text{h.c.}, \quad c_{4\alpha}^\dagger c_{5\alpha} c_{6\beta} c_{7\beta}^\dagger + \text{h.c.}\}.\end{aligned} \tag{S42}$$

$$\begin{aligned}Q_{C_6H_6} = \{&c_{2\alpha}^\dagger c_{4\alpha}^\dagger c_{4\alpha} c_{5\alpha} + \text{h.c.}, \quad c_{2\alpha}^\dagger c_{4\alpha} c_{4\alpha} c_{5\alpha}^\dagger + \text{h.c.}, \quad c_{2\beta}^\dagger c_{2\beta}^\dagger c_{4\beta} c_{4\beta} + \text{h.c.},\\ &c_{2\beta}^\dagger c_{2\beta} c_{4\beta} c_{4\beta}^\dagger + \text{h.c.}, \quad c_{2\alpha}^\dagger c_{4\alpha} c_{4\beta}^\dagger c_{5\beta} + \text{h.c.}, \quad c_{1\alpha}^\dagger c_{3\alpha} c_{5\beta} c_{5\beta}^\dagger + \text{h.c.}\}.\end{aligned} \tag{S43}$$

$$\begin{aligned}Q_{FeCp_2} = \{&c_{3\alpha}^\dagger c_{6\alpha}^\dagger c_{7\alpha} c_{7\alpha} + \text{h.c.}, \quad c_{3\alpha}^\dagger c_{6\alpha} c_{7\alpha} c_{7\alpha}^\dagger + \text{h.c.}, \quad c_{3\beta}^\dagger c_{6\beta}^\dagger c_{7\beta} c_{7\beta} + \text{h.c.},\\ &c_{3\beta}^\dagger c_{6\beta} c_{7\beta} c_{7\beta}^\dagger + \text{h.c.}, \quad c_{3\alpha}^\dagger c_{6\alpha} c_{7\beta}^\dagger c_{7\beta} + \text{h.c.}, \quad c_{3\alpha}^\dagger c_{6\alpha} c_{7\beta} c_{7\beta}^\dagger + \text{h.c.}\}.\end{aligned} \tag{S44}$$

## S10 Detailed numerical results for atomic spectra

In this section, we summarize the detailed numerical results for all the atomic systems considered in this work in Table S1 below. These results correspond to those presented in Sec. 2.4 in the main text. The table includes information about the atomic system, the electronic state being targeted, the total Lindbladian simulation time, the final energy in Hartree (Ha), and the final absolute error in energy compared to the FCI solution, the infidelity of the final state with respect to the exact eigenstate obtained by FCI calculations, the Lindblad simulation time taken to reach chemical accuracy, and the final expectation value of $2S + 1$ indicating the spin multiplicity of the prepared state. The system is deemed to have reached chemical accuracy at the earliest time point for which the absolute energy error falls below 1.6 mHa and remains below this threshold for the subsequent 20 time steps.

Table S1: Detailed numerical results for atomic spectra

| System | State | Simulation time | Coupling operators | Energy (Ha) | Error (Ha) | Infidelity | Time to reach chem. acc. | $2S + 1$ |
|---|---|---|---|---|---|---|---|---|
| Li | $^2P(2p^1)$ | 30 | $S'_{II}$ | −7.23048 | $3.02274 \times 10^{-11}$ | $4.44189 \times 10^{-12}$ | 2.01 | 2.000 |
| Be | $^3P(2s^22p^1)$ | 30 | $S'_{II}$ | −14.2866 | $5.53513 \times 10^{-12}$ | $5.60463 \times 10^{-12}$ | 2.62 | 3.000 |
| Be | $^1P(2s^12p^1)$ | 30 | $S'_{II}$ | −14.1137 | $1.76634 \times 10^{-9}$ | $6.55749 \times 10^{-8}$ | 3.22 | 1.000 |
| B | $^4P(2s^12p^2)$ | 30 | $S'_{II}$ | −24.0756 | $4.03446 \times 10^{-11}$ | $1.80939 \times 10^{-11}$ | 4.51 | 4.000 |
| C[a] | $^1D(2s^22p^2)$ | 30 | $S'_{II} \cup Q$ | −37.1462 | $6.21725 \times 10^{-12}$ | $1.79786 \times 10^{-11}$ | 1.40 | 1.000 |
| C[a] | $^5S(2s^12p^3)$ | 50 | $S'_{II} \cup Q$ | −37.1090 | $1.33238 \times 10^{-5}$ | $8.33338 \times 10^{-4}$ | 8.52 | 4.998 |
| C[a] | $^1S(2s^22p^2)$ | 30 | $S'_{II} \cup Q$ | −37.0934 | $7.67386 \times 10^{-13}$ | $5.77316 \times 10^{-15}$ | 1.50 | 1.000 |
| N | $^2D(2s^22p^3)$ | 30 | $S'_{II}$ | −53.5957 | $1.35915 \times 10^{-7}$ | $1.10131 \times 10^{-6}$ | 0.91 | 2.000 |
| N | $^2P(2s^22p^3)$ | 30 | $S'_{II}$ | −53.5529 | $1.10724 \times 10^{-10}$ | $2.74170 \times 10^{-13}$ | 1.21 | 2.000 |
| O | $^1D(2s^22p^4)$ | 30 | $S'_{II}$ | −73.7093 | $4.80108 \times 10^{-6}$ | $5.79038 \times 10^{-5}$ | 0.60 | 1.000 |
| O | $^1S(2s^22p^4)$ | 30 | $S'_{II}$ | −73.6274 | $4.72369 \times 10^{-11}$ | $4.48530 \times 10^{-14}$ | 1.30 | 1.000 |

[a] For the carbon atom, in the complete basis set limit, the eigenstate ordering is $^3P(2s^22p^2) < {}^1D(2s^22p^2) < {}^1S(2s^22p^2) < {}^5S(2s^12p^3)$ (6). The discrepancy observed here is due to the finite basis set effects.

## S11 Detailed numerical results for molecular systems

In this section, we summarize the detailed numerical results for all the molecular systems considered in this work in Table S2 below. These results correspond to those presented in Sec. 2.5 and Sec. 2.6 in the main text. Again, the benchmark exact solutions are obtained by FCI or CASCI exact diagonalization calculations. The criteria for reaching chemical accuracy is the same as that in Table S1.

## S12 Geometries of molecular systems

The geometries of the molecular systems considered in this work are provided below. Distances are given in Angstroms (Å).

Table S2: Detailed numerical results for different molecular systems.

| System | State | Simulation time | Coupling operators | Energy (Ha) | Error (Ha) | Infidelity | Time to reach chem. acc. |
|---|---|---|---|---|---|---|---|
| BH | $^3\Pi$ | 20 | $S'_{\text{II}}$ | −24.6563 | $3.68496 \times 10^{-9}$ | $7.30039 \times 10^{-9}$ | 3.03 |
|  | $^1\Pi$ | 20 | $S'_{\text{II}}$ | −24.6535 | $3.59055 \times 10^{-10}$ | $1.32825 \times 10^{-9}$ | 2.63 |
| CH$^+$ | $^3\Pi$ | 20 | $S'_{\text{II}}$ | −37.3902 | $5.62423 \times 10^{-10}$ | $7.22172 \times 10^{-10}$ | 2.42 |
|  | $^1\Pi$ | 20 | $S'_{\text{II}}$ | −37.3830 | $4.16343 \times 10^{-10}$ | $5.34812 \times 10^{-10}$ | 2.63 |
| HCl | $^3\Pi$ | 20 | $S'_{\text{II}}$ | −454.616 | $3.69482 \times 10^{-12}$ | $1.69755 \times 10^{-13}$ | 2.04 |
|  | $^1\Pi$ | 20 | $S'_{\text{II}}$ | −454.578 | $1.98952 \times 10^{-12}$ | $1.11022 \times 10^{-16}$ | 1.43 |
| H$_2$O[a] | $^3B_1$ | 20 | $S'_{\text{II}}$ | −74.6146 | $1.05160 \times 10^{-12}$ | $5.77316 \times 10^{-15}$ | 5.66 |
|  | $^1B_1$ | 20 | $S'_{\text{II}}$ | −74.5549 | $5.68434 \times 10^{-14}$ | $4.62963 \times 10^{-14}$ | 6.67 |
|  | $^3A_1$ | 20 | $S'_{\text{II}} \cup Q$ | −74.5110 | $1.44951 \times 10^{-12}$ | $3.33067 \times 10^{-16}$ | 2.63 |
|  | $^3A_2$ | 20 | $S'_{\text{II}} \cup Q$ | −74.5088 | $4.77257 \times 10^{-6}$ | $7.44078 \times 10^{-6}$ | 2.02 |
|  | $^1A_2$ | 20 | $S'_{\text{II}}$ | −74.4715 | $1.04876 \times 10^{-11}$ | $1.03881 \times 10^{-11}$ | 4.24 |
| C$_6$H$_6$[b] | $1^1B_{2u}$ | 20 | $S'_{\text{II}}$ | −230.6647 | $1.77636 \times 10^{-15}$ | $8.88178 \times 10^{-16}$ | 2.31 |
|  | $1^1B_{1u}$ | 20 | $S'_{\text{II}} \cup Q$ | −230.5566 | $3.79259 \times 10^{-5}$ | $3.33382 \times 10^{-3}$ | 6.83 |
|  | $2^1E_{2g}$ | 50 | $S'_{\text{II}} \cup Q$ | −230.5456 | $3.54364 \times 10^{-6}$ | $8.26061 \times 10^{-4}$ | 0.80 |
|  | $1^1E_{1u}$ | 50 | $S'_{\text{II}} \cup Q$ | −230.5052 | $5.91244 \times 10^{-5}$ | $2.01584 \times 10^{-3}$ | 9.32 |
| FeCp$_2$[c] | $1^3E''_1$ | 50 | $S'_{\text{II}}$ | −1655.9239 | $8.61462 \times 10^{-6}$ | $4.44587 \times 10^{-4}$ | 1.51 |
|  | $1^3E''_2$ | 50 | $S'_{\text{II}}$ | −1655.9230 | $2.42764 \times 10^{-10}$ | $2.09724 \times 10^{-10}$ | 1.21 |
|  | $1^1E''_2$ | 50 | $S'_{\text{II}}$ | −1655.8859 | $5.65556 \times 10^{-11}$ | $3.75777 \times 10^{-12}$ | 2.61 |
|  | $1^1E''_1$ | 50 | $S'_{\text{II}}$ | −1655.8679 | $3.87322 \times 10^{-6}$ | $3.20185 \times 10^{-4}$ | 1.01 |
|  | $2^3E''_1$ | 50 | $S'_{\text{II}} \cup Q$ | −1655.8339 | $2.30109 \times 10^{-8}$ | $9.72585 \times 10^{-7}$ | 3.62 |
|  | $2^1E''_1$ | 100 | $S'_{\text{II}} \cup Q$ | −1655.7765 | $2.09317 \times 10^{-6}$ | $2.66667 \times 10^{-3}$ | 1.10 |

[a] For the H$_2$O example, in the complete basis set limit, the eigenstate ordering is $^1A_1 < {}^3B_1 < {}^1B_1 < {}^3A_2 < {}^1A_2 < {}^3A_1$ (6). The discrepancy observed here is due to the finite basis set effects.

[b] For the C$_6$H$_6$ example, the experimental ordering of the singlet excited states is $^1B_{2u} < {}^1B_{1u} < {}^1E_{1u} < {}^2E_{2g}$ (48, 49). The discrepancy observed here is due to the fact that we only include the valence orbitals in our (6e, 6o) active space modeland therefore neglects dynamical correlation effects. See detailed discussion in Sec. 2.6 in the main text.

[c] For the ferrocene example, the experimental ordering of the low-lying excited states is $1^3E''_1 < 1^3E''_2 < 2^3E''_1 < 1^1E''_2 < 1^1E''_1 < 2^1E''_1$ (50, 54). The discrepancy observed here is due to the fact that we only include the valence orbitals in our (10e, 7o) active space model. See detailed discussion in Sec. 2.6 in the main text.

Table S3: **Geometry of BH molecule**

| Atom | x | y | z |
|------|-----|-----|--------|
| B | 0.0 | 0.0 | 1.243 |
| H | 0.0 | 0.0 | −1.243 |

Table S4: **Geometry of CH$^+$ molecule**

| Atom | x | y | z |
|------|-----|-----|--------|
| C | 0.0 | 0.0 | 1.131 |
| H | 0.0 | 0.0 | −1.131 |

Table S5: **Geometry of H$_2$O molecule**

| Atom | x | y | z |
|------|--------|---------|---------|
| O | 0.0000 | 0.0000 | 0.1173 |
| H | 0.0000 | 0.7572 | −0.4692 |
| H | 0.0000 | −0.7572 | −0.4692 |

Table S6: **Geometry of C$_6$H$_6$ molecule**. See Supplementary Note S7 and Ref. (*40*).

| Atom | x | y | z |
|------|-----------|-----------|----------|
| C | 0.000000 | 0.001262 | 0.000000 |
| C | 0.000000 | 2.782738 | 0.000000 |
| C | 1.204419 | 0.696631 | 0.000000 |
| C | −1.204419 | 0.696631 | 0.000000 |
| C | 1.204419 | 2.087369 | 0.000000 |
| C | −1.204419 | 2.087369 | 0.000000 |
| H | −2.141507 | 2.628401 | 0.000000 |
| H | 2.141507 | 2.628401 | 0.000000 |
| H | −2.141507 | 0.155599 | 0.000000 |
| H | 2.141507 | 0.155599 | 0.000000 |
| H | 0.000000 | −1.080796 | 0.000000 |
| H | 0.000000 | 3.864796 | 0.000000 |

Table S7: **Geometry of ferrocene (FeCp$_2$) molecule**. The geometry is taken from Ref. (77), see also Supplementary Note S7 and Ref. (39).

| Atom | x | y | z |
|------|-----------|-----------|-----------|
| Fe | 0.000000 | 0.000000 | 0.000000 |
| C | −0.713500 | −0.982049 | −1.648000 |
| C | 0.713500 | −0.982049 | −1.648000 |
| C | 1.154467 | 0.375109 | −1.648000 |
| C | 0.000000 | 1.213879 | −1.648000 |
| C | −1.154467 | 0.375109 | −1.648000 |
| H | −1.347694 | −1.854942 | −1.638208 |
| H | 1.347694 | −1.854942 | −1.638208 |
| H | 2.180615 | 0.708525 | −1.638208 |
| H | 0.000000 | 2.292835 | −1.638208 |
| H | −2.180615 | 0.708525 | −1.638208 |
| C | −0.713500 | −0.982049 | 1.648000 |
| C | −1.154467 | 0.375109 | 1.648000 |
| C | −0.000000 | 1.213879 | 1.648000 |
| C | 1.154467 | 0.375109 | 1.648000 |
| C | 0.713500 | −0.982049 | 1.648000 |
| H | −1.347694 | −1.854942 | 1.638208 |
| H | −2.180615 | 0.708525 | 1.638208 |
| H | 0.000000 | 2.292835 | 1.638208 |
| H | 2.180615 | 0.708525 | 1.638208 |
| H | 1.347694 | −1.854942 | 1.638208 |



\end{document}